\documentclass[aps,prb,twocolumn,groupedaddress,showpacs,showkeys]{revtex4-1}

\usepackage{amsmath,amssymb}
\usepackage{graphicx} 
\usepackage{dcolumn}
\usepackage{bm}
\usepackage{times}
\usepackage{longtable} 
\usepackage{hyperref} 

\graphicspath{ {/home/accancio/Work/exc_fit/ExqlaplPaper/}
               {/home/accancio/Work/exc_files/exlapl2.0/work/Umrigar} }

\newcommand{\be}{\begin{equation}}
\newcommand{\ee}{\end{equation}}
\newcommand{\bea}{\begin{eqnarray}}
\newcommand{\eea}{\end{eqnarray}}
\newcommand{\beas}{\begin{eqnarray*}}
\newcommand{\eeas}{\end{eqnarray*}}

\newcommand{\bfs}{{\bf s}}

\newcommand{\bfr}{{\bf r}}
\newcommand{\bfv}{{\bf v}}
\newcommand{\bfrp}{{\bf r}^\prime}

\newcommand{\bfu}{{\bf u}}

\newcommand{\Exc}{ E_{XC} }

\newcommand{\Fx}{F_X}

\newcommand{\exc}{e_{XC}}
\newcommand{\epsxc}{ \epsilon_{XC} }
\newcommand{\epsx}{ \epsilon_{X} }
\newcommand{\nxc}{ n_{XC} }

\newcommand{\lapln}{\nabla^2n}

\newcommand{\gradn}{\nabla n}

\newcommand{\gradnsq}{\left|\nabla n\right|^2}

\newcommand{\dexcdlapln}{\frac{\partial \exc}{\partial \lapln}}

\newcommand{\xbar}{\bar{x}}
\newcommand{\mubar}{\bar{\mu}}
\newcommand{\barx}{\bar{x}}

\begin{document}

\preprint{Cancio and Wagner, preprint 2013}

\title{
Laplacian-based generalized gradient approximations for the exchange energy
}


\author{Antonio C. Cancio}
\affiliation{Department of Physics and Astronomy, Ball State University, Muncie, IN 47306}
\email[]{accancio@bsu.edu}
\author{Chris E. Wagner}
\affiliation{Department of Physics, 
University of Florida, Gainesville, FL 32611}


\date{\today}

\begin{abstract}
It is well known that in the gradient expansion approximation to density functional theory (DFT) the gradient and Laplacian of the density make interchangeable contributions to the exchange correlation (XC) energy.  This is an arbitrary ``gauge" freedom for building DFT models, normally used to eliminate the Laplacian from the generalized gradient approximation (GGA) level of DFT development.  We explore the implications of keeping the Laplacian at this level of DFT, to develop a model that fits the known behavior of the XC hole, which can only be described as a system average in conventional GGA.  We generate a family of exchange models that obey the same constraints as conventional GGA's, but which in addition have a finite-valued potential at the atomic nucleus unlike GGA's.  These are tested against exact densities and exchange potentials for small atoms, and for constraints chosen to reproduce the SOGGA and the APBE variants of the GGA.  The model reliably reproduces exchange energies of closed shell atoms, once constraints such the local Lieb-Oxford bound, whose effects depend upon choice of energy-density gauge, are recast in invariant form.  
\end{abstract}


\maketitle


\section{Introduction}
\subsection{The exchange-correlation energy and hole}
The basic issue of density functional theory~\cite[]{HK,DreizlerGross,JG} (DFT) 
is modeling the 
exchange-correlation (XC) energy -- the
description of the electron-electron interaction energy due to
Fermi statistics and many-body interactions within a single-particle formulation
of the ground-state problem.  Beyond the very basic local
density approximation (LDA) which maps the 
XC energy-per-particle at any point in space to that of the
homogeneous electron gas (HEG),~\cite{KohnSham}
there is no truly systematic approach to building functionals.
A heuristic paradigm that provides some structure to the problem is 
that of a  Jacob's ladder~\cite{JacobsLadder}
of functionals, where these are grouped in families (rungs on the 
ladder) with each successive rung characterized by an increase in the 
amount of information about the system being used.
Generally, this leads to functionals of increasing sophistication
and accuracy as one moves up the ladder, 
but with the tradeoff of increasing complexity and
computational cost.  

Left unsaid is how one is to obtain the best description of
nature within a given class or rung of functionals.  
A good example of this lack of systematizability within a class 
is the low-level generalized gradient approximation
(GGA) family of functionals that employ the gradient of the density as
well as the local value of the density to construct local XC energies.
This class of functional is extremely popular, easy to implement and use, 
but not quite robust enough for enough applications to be a 
completely satisfactory workhorse for electronic structure calculations. 
There has thus been much effort to
optimize the GGA and dozens of published functionals.  Taking one
commonly used functional, the PBE,~\cite{PBE} variants have been
developed to optimize it either for specific applications such as 
atomization energies for molecules,~\cite{RPBE,revPBE,APBE,PBEmol}
structural constants, particularly of solids,~\cite{WuCohen, SOGGA} 
or solid surface energies,~\cite{PBEsol} 
or for broad applicability.~\cite{RGE2,PBEint,SystematicCapelle}  
It has been argued that a GGA in principle does not 
include enough information to reach all these goals.~\cite{PCSB-PRL97,revTPSS}

A particularly fruitful approach to understanding 
the XC energy has been the XC hole, $n_{xc}(\bfr,\bfr^{\prime})$,
defined as the change in electron density 
from the mean at $\bfr^{\prime}$ given the
presence of an electron observed to be at $\bfr$. 
        %
       %
      An exchange-correlation energy-per-particle $\epsxc$ at any point
      $\bfr$   
      can then be defined 
      as the net change in energy due to the formation of the hole
      about an electron placed at $\bfr$.
      This 
      includes the potential energy gain 
      due to the interaction of the electron with its own hole and
      the kinetic energy cost to create it.  The net effect is
      obtained through an adiabatic 
      formulation\cite[]{HarrisJones,LPadiabatic,GunnLund}
    \be
      \label{eq:epsxcadiabatic}
        \epsxc(\bfr) = \frac{1}{2} \int d\lambda\;
                        \int_0^1 d^3r^\prime\;
                           \frac{ \nxc^{\lambda}(\bfr,\bfr^{\prime}) }
                                { \left| \bfr - \bfr^{\prime}  \right| }.
    \ee
(Throughout this paper, expressions are written
in hartree atomic units.)
Here $\nxc^{\lambda}$ is the XC hole evaluated for
a system with Coulomb coupling $\lambda e^2$ and the same ground-state
density $n(\bfr)$ as the true system.  The coupling constant varies from 
noninteracting (0) to fully interacting (1) systems. 
The total XC energy is obtained from $\epsxc$ by integration over $\bfr$:
      \be
         E_{xc} = \int d^3r\; \exc(\bfr) = \int d^3r\; n(\bfr) \epsxc(\bfr).
         \label{eq:exc}
      \ee
      where $\exc(\bfr) = n(\bfr)\epsxc(\bfr)$ is the local XC energy
      density. 
The LDA can then be derived by parametrizing $\epsxc(\bfr)$ in terms of the
local density $n(\bfr)$ and the GGA by both local density and 
gradient $\gradn(\bfr)$.

      The close connection between the XC hole 
      and energy 
      historically provided a theoretical basis to explain 
      the surprising success\cite{JG} of the LDA and to
      to construct widely-used DFT's
      including the PBE,\cite[]{PBW_GGA} and hybrids with
      Hartree-Fock.\cite[]{BeckeHybrid}  The XC hole has played 
      a supporting role in recent developments
      in modeling the XC potential,~\cite[]{BeckeJohnson,BeckeRoussel}
      as well as sophisticated ``Koopman's compliant" functionals to treat 
      many-electron self-interaction error;~\cite{MarzariKoopmans} the adiabatic connection
      also continues to be of importance in a variety of contexts.~\cite{CohenMoriYang}

\subsection{Revisiting the gradient expansion}

A major problem with using the energy density $\exc$ to define a 
XC functional is that it is not uniquely defined.  
    Adding any functional of the density that
    integrates to zero for any density will define a new $\exc$ while leaving the 
    XC energy, and thus the physics of the system, unchanged.  
    This is not hard to do -- for example, the addition of a divergence of
    a vector functional is enough:
    \be
         \exc' = \exc[n](\bfr) + \nabla \cdot \bfv[n](\bfr),
    \ee
    as long as the related flux integral can be set to zero on the boundary.
    If such a relation between $\exc$ and $\exc'$ holds
     for every possible density $n(\bfr)$, then the total energy and also 
    the potential $\delta \Exc/\delta n(\bfr)$ of each 
    will be indistinguishable from each other. 
    This ambiguity is thus similar to that of the gauge of the 
    potential in classical electrodynamics.

The gauge ambiguity in defining $\exc$ played a role 
from the very beginning of DFT with the 
gradient expansion.~\cite{HK}  Hohenberg and Kohn, in their seminal paper on DFT
note that the gradient expansion for a system with a slowly varying density 
should most generally be expressed as 
   \be
     e^{GEA}_{XC}[n] = e_{XC}^0[n] + \left \{
               e_{XC}^{(2a)}[n] \nabla^2 n + 
               e_{XC}^{(2b)}[n] | \nabla n |^2
        \right\}   + O(\nabla^4) 
   \label{eq:GEA}
   \ee
with two second order terms, involving the gradient and the Laplacian of the 
density.
With the addition of the pure divergence $\nabla \cdot (-\exc^{(2a)}\nabla n)$,
the energy density can be converted into a form exclusively involving the
gradient-squared of the density, obviating the need of a Laplacian term
entirely and making life simpler for modelers. 

Nevertheless, there are reasons for exploring the gauge freedom to 
construct DFT from different starting points.  It has been a long standing
issue of DFT's that build upon the gradient-only form of the GEA that 
the XC potentials they produce have 
a false $1/r$ singularity in the XC potential in the vicinity
of the atomic nucleus, induced by the cusp in the charge density in this limit.
However, it is easy~\cite{UmrigarGonze} to construct a finite potential if one has
access to the Laplacian. 


    Secondly, by transforming $\epsxc$ to a gradient-only
    form, the fruitful connection between XC energy and XC hole becomes 
    obscured.  We show below that the local energy of the X hole in the gradient
    expansion limit is necessarily a functional of both the gradient and the Laplacian; 
    although an accurate local energy is not needed to get correct total exchange
    energy, it helps to have a correct local energy to analyze 
    trouble areas like the 1s shell and nuclear cusp.
Moreover, by eliminating the Laplacian of the density, one 
is discarding information about the topology of an electronic 
system that is potentially useful, especially if 
tied back to the XC hole.  For example, the Laplacian is known to be 
a faithful indicator of shell structure and useful in
diagnosing the ionicity of bonds, used for this
purpose in the Atoms in Molecules approach to visualizing
molecular structure and reactions.~\citep{Bader1,Bader2}
  
  This issue has been highlighted by recent work modeling data for the 
exchange-correlation hole
from variational quantum Monte Carlo (VMC) studies.~\cite{CancioLapl}
          Recent simulations have obtained data for 
          the adiabatically integrated XC hole and
          energy density using highly
          accurate VMC methods for calculating the
          expectations of an optimized many-body wavefunction
           and explicit coupling-constant integration.\cite[]{Hood2}
          These include the Si crystal,\cite[]{Hood1,Hood2,Cancio} 
          atoms,\cite[]{CancioFong,Puzder,Cancio2RowPaper} and small 
       organic molecules\cite[]{Hsing} within a pseudopotential approximation,
       and a model charge-density-wave system.\cite[]{Nekovee1,Nekovee2}
The outstanding feature of all these studies
is the strong correlation between the local Laplacian of the density and 
the error in the LDA model for the adiabatically integrated XC energy density,
measured with respect to the VMC data.
   The correlation is unmistakable and
   cannot be described in terms of alternate variables such as the
       kinetic energy density or the gradient of the density.  
    An empirical fit of this difference for the Si crystal 
    to a Laplacian-based enhancement factor of the LDA energy, 
    constructed in analogy to a GGA, 
    recovers 70\% of the energy difference between the VMC and the LDA for
    most of the systems studied.\cite{CancioLapl}

       Thus VMC data indicate that possibly we should rethink the
       GGA ``rung" in the Jacob's ladder for DFT.  
       Apparently nature, or at least the adiabatic XC hole is better described
       using a different ``gauge" choice than that of the classic GGA, and 
       designing a GGA based on such a gauge may help us to use the insights from
       XC hole calculations more effectively.
       And the extra degree of freedom allows for applying constraints that are 
       not possible in the classic GGA, such as the character of the XC potential
       at the nucleus.
       It is possible also that redefinition of the GGA in terms
       of the Laplacian might lead to 
       better ground state predictions,
       such as that of covalently bonded systems such as the Si crystal.

       The Laplacian of the density has not been used much
       in DFT, with but a few explicit models~\cite{PerdewConstantin} 
       in recent years.  One approach to DFT that does employ them
       is complementary to one taken here -- use
       of X and C holes to define corresponding potentials, with a fit to
       the expansion of the exact exchange hole to determine shape and
       extent of hole.~\cite{BeckeRoussel}  
       Laplacian terms naturally arise there for the same reason as they
       do in this paper, from the description of the exchange hole, while
       the approach is computationally quite a bit more intensive than
       a GGA, and belongs on the higher ``metaGGA" class of 
       functionals.  This approach has had a revival
       in recent years,~\cite{BeckeJohnson,Rasanen,TranBlaha} producing
       very accurate potentials  and improved semiconductor band-gaps.
       However the path back to extracting useful XC energies from potentials
       is nontrivial.~\cite{Gaiduk}

Finally, we note the relevance of this work 
to the development of  orbital-free DFT's. 
Explicit functionals of the density, eliminating the 
step of constructing orbitals as must be done for the kinetic energy
density in the Kohn-Sham approach, would be of great usefulness
for extending the capability of DFT. This is especially true for application
to warm dense matter where both the conditions of quantum
mechanics and high temperature (thus high orbital occupancies) must 
be dealt with.~\cite{KarasievOFDFT}
Laplacian-based XC functionals are of interest in this context as potential
candidates to replace meta-GGA's which currently rely upon the 
Kohn-Sham kinetic energy density, and thus require the use of orbitals.
And the techniques developed here for implementing this variable in the XC 
functional may be of use for KE functionals as well.

The focus of this paper is on designing a mature, robust functional
for exchange.  
Correlation is, to a large degree, a response to exchange, and as 
a result, parameters and functional forms for correlation are 
chosen to fit with the given form for exchange.  
At the same time, correlation tends to be considerably more 
complex than exchange because it lacks the simple scaling behavior
under uniform coordinate scaling that restricts exchange to 
a relatively simple form.  Nevertheless many of the techniques discussed
here should be applicable to correlation as well.

In a preliminary attempt at the design of a Laplacian-based GGA for 
exchange,~\cite{CancioExlapl} 
we demonstrated the possibility of using the Laplacian in combination
with the gradient of the density in density functional theory 
in an all-electron context, going beyond the pseudopotential 
approximations used in QMC.
We focused on resolving technical issues of controlling nonlinear behavior in 
the exchange-correlation potential due to the use of Laplacian and
poor choices for the form of the functional.
In this paper, we develop and construct an 
effective set of ``gauge invariant" constraints 
based on those of PBE, requiring us to revisit especially the bounds
in the limit of large inhomogeneity.  
The use of the Laplacian allows (and requires) one 
to satisfy a number of constraints on the potential in addition to the 
energy, which results in a more robust functional.   
We test our functional against numerically exact densities 
and potentials for small atoms and against densities 
and energies derived from the APBE functional~\cite{APBE} for large atoms.  
The end result is an effective and mature exchange functional that 
reproduces GGA energies for atoms and fixes the problem of the GGA 
potential at the nucleus.

Section~\ref{sec:theory} discusses in detail the theory behind our
functional, especially the constraints used; Section~\ref{sec:methods} is
a short description of numerical techniques used in testing our functionals.
Energies and potentials for example atoms are discussed 
in Section~\ref{sec:results} and
Section~\ref{sec:discussion} includes a discussion of possible future
steps and our conclusions.

\section{Theory \label{sec:theory}}
We start with the basic form of the PBE functional, perhaps the
    most commonly used constraint-based GGA and the generator of 
    a large family of derivative functionals.
    Ignoring spin polarization, the PBE energy density is given by:
    \be  
    \begin{split}
     e^{PBE}_{xc}(n, \nabla n) = 
              F^{PBE}_X(s^2) e_X^{LDA}(r_s)\\  + \:
              e_C^{LDA} (r_s) + H^{PBE}_C(r_s,t^2).
    \label{eq:PBE}
    \end{split}
    \ee
    where the two basic order parameters characterizing the energy are
    the Wigner-Seitz radius $r_s \!=\! {3/4\pi n}^{1/3}$, and a scale-invariant
    inhomogeneity parameter,
    \be
        s^2 = \frac{| \nabla n |^2 } {4k_F^2 n^2},
        \label{eq:stwo}
    \ee
    with $k_F \!=\! (3\pi^2 n)^{1/3} \!\sim\! 1/r_s$ being the local Fermi wavevector.
The LDA exchange energy density is $e_x^{LDA}\!=\! -3k_Fn/4\pi$ and scales with 
density as $n^{4/3}$.  It shows the correct behavior under an important
scaling transformation -- the uniform scaling of coordinates
$r \!\rightarrow\! \gamma r$, $n(r) \!\rightarrow\! \gamma^3 n(\gamma r)$. 
To preserve this behavior, the GGA correction is restricted
in form to   
a multiplicative enhancement factor $F_X$ parameterized solely in terms of 
scale-invariant quantities such as $s^2$.

The LDA correlation energy density 
is a function of $r_s$ with non-trivial scaling behavior.
The correlation correction $H_C$ is, likewise, a 
more complex functional than its exchange counterpart, depending on a 
inhomogeneity parameter $t^2$ that is defined
with respect to  the Thomas-Fermi screening vector
$k_s\!=\!\sqrt{4k_F/\pi a_0}$.  This does not introduce a new 
variable to the functional, as it reduces to a function of $s^2$
and $k_F$:
\be
    t^2 = (k_F/4\pi)s^2 
    \label{eq:tsq}
\ee
The additive form of the generalized gradient correction for correlation, $H_C$,
is, like the multiplicative form $F_X$ for exchange, 
necessitated by the properties of correlation under uniform 
coordinate scaling.
Finally, the details of the functional form of $F_X$ and
$H_C$ are adapted so as to satisfy other known constraints, particularly in the 
slowly varying limit $s^2 \!\ll\! 1$ and in the limit of extreme inhomogeneity
$s^2 \!\gg\! 1$.  

We now reengineer the PBE to generate the equivalent functional based upon
an arbitrary linear combination of gradient and Laplacian of the density along
the lines of the most general gradient expansion form, Eq.~(\ref{eq:GEA}).
In the slowly-varying limit appropriate for the gradient expansion, 
$s^2 \!\rightarrow\! 0$, the spin-unpolarized PBE reduces to 
\be
    \exc^{GEA} = [1 + \mu s^2 ] e_x^{LDA}( r_s ) + e_c^{LDA}(r_s) - n \beta t^2 
\ee
where $\mu \!=\! 10/81$ and $\beta \!=\! 0.066725$ determine the strength of the
gradient correction and can be obtained from perturbation 
theory.\cite{Svendsen,MaBrueckner}
A key point to note is that the correlation gradient
correction $n t^2$ has the same functional form as that for
exchange -- both vary with the density as $n^{-4/3} \gradnsq$.  
The overall XC gradient correction can thus be recast explicitly in the
form of Eq.~(\ref{eq:GEA}):
\be
    \exc^{GEA}-\exc^{LDA} = (\mu\! -\! \mu_c) s^2 e_x^{LDA}( r_s ) \sim n^{-4/3}\gradnsq
    \label{eq:PBE-GEA}
\ee
with $\mu_c \!=\! \beta (\pi^2/3)$. 
Correlation has the effect of reducing the overall correction to the LDA 
-- in a sense, departures from the LDA exchange
hole due to inhomogeneity tend to induce a compensating response in the 
correlation hole.
Numerical data suggest~\cite{Moroni,Ortiz} 
that in the limit of small perturbations of the HEG, i.e. linear response, 
this compensation is perfect: $\mu\! = \!\mu_c$.  Unfortunately this 
is not consistent with the values of $\mu$ and $\beta$ obtained from perturbation
theory.

Eq.~(\ref{eq:PBE-GEA}) describes the gradient expansion using an energy 
density in the ``gauge" that eliminates the term proportional to $\lapln$ 
that appears in the more general form. 
If we were to reverse this process, 
by an integration by parts leaving the total XC energy
unchanged,  
we come up with a Laplacian-only GEA of the form:~\cite{FootnoteThree}
\be
    \exc^{GEA}\!-\!\exc^{LDA} = (\mu \!- \!\mu_c) \;3q\; e_x^{LDA}( r_s ) \sim n^{-1/3}\lapln
    \label{eq:PBE-qGEA}
\ee
where $q$ is given by
    \be
        q = \frac{\nabla^2 n} {4k_F^2 n}. 
        \label{eq:q}
    \ee
And more generally, we can consider a generalized gradient expansion
variable $x$, a hybrid of $q$ and $s^2$, defined in terms of a gauge
parameter $\alpha$ that can be continuously varied between 0 ($s^2$ only)
and 1 ($q$ only):
\be
    x = (1-\alpha)s^2/3 + \alpha q
    \label{eq:x}.
\ee
One can then generate an entire family of GGA's by 
replacing $s^2$ with $3x$ in Eq.~(\ref{eq:PBE}) and Eq.~(\ref{eq:PBE-GEA}),
changing the parameter
$\alpha$ while keeping the basic form of the GGA fixed.
The generalized PBE form becomes
    \be  
    \begin{split}
     e^{modPBE}_{xc}(n, \nabla n) = 
              F^{PBE}_X(3x) e_X^{LDA}(r_s) \\
              + \: e_C^{LDA} (r_s) + H^{PBE}_C(r_s,3y),
    \label{eq:xPBE}
    \end{split}
    \ee
with $y \!=\! x(t^2/s^2)$ being the most general gradient-expansion 
inhomogeneity parameter for correlation.
One thus has a new ``dial" for manipulating the GGA
to improve the robustness of the functional in regions and systems of 
high inhomogeneity while keeping the gradient expansion correction for
slowly varying systems unaltered.

A final issue of importance to the development of a Laplacian-based
functional is how the Laplacian affects the exchange-correlation potential,
$V_{XC}$.
The exchange-correlation potential for a functional that depends 
explicitly on the local density, its gradient and Laplacian is given by
\be
V_{XC} = \frac{\partial e_{XC} } {\partial n} 
     - \nabla \cdot \left(  \frac{\partial e_{XC}} {\partial \nabla n}
                   \right) 
     + \nabla^2 \left ( \frac{\partial e_{XC}} {\partial \nabla^2 n}\right ).
\label{eq:vxc}
\ee
The second term comes from the variation of $E_{XC}$ with $\gradn$ and the
third from the variation with respect to $\lapln$ and does not appear in
a normal GGA.  Both derivative terms can be causes of difficulty in
DFT development -- the divergence operator in the second term 
causes a $1/r$ singularity at the nucleus for GGA's while the large number
of derivatives (up to $\nabla^4 n$) can be a cause of instability 
in the potential for Laplacian-based functionals.

\subsection{Choice of gauge parameter $\alpha$}

Given the strong correlation seen in QMC data between the energy density 
associated with the standard definition of the XC hole 
[Eq.~(\ref{eq:epsxcadiabatic})], it seems 
that nature ``prefers" a gauge choice with nonzero $\alpha$.  
But what choice of $\alpha$ best matches the energy density constructed
from the adiabatic XC hole?  A natural guess is to try to construct a 
gradient expansion of the exchange-correlation hole and derive a value of 
$\alpha$ from this. 

This strategy can in part be carried out using an expansion of the 
exchange hole only
derived by Becke~\citep{Beckenxc,BeckemGGA} in developing an early metaGGA.  
He expanded the well-known analytic expression for the exchange hole at small 
electron-electron separation $u$ to second order in $u$ to find
\be
\begin{split}
   \langle n_{X}(\bfr, u)\rangle {}& = \frac{n(\bfr)}{2} \: + \\
         {}& \frac{1}{12} 
         \left[ \nabla^2 n(\bfr) - 4\tau(\bfr) + \frac{1}{2} \sum_\sigma
                         \frac{\left|\nabla n_\sigma(\bfr)\right|^2}{n_\sigma(\bfr)}
         \right] u^2
    \label{eq:expandnx}
\end{split}
\ee
Here the brackets indicate a spherical average over the angle $\Omega_u$ 
of the interparticle displacement $\bfu$, 
and $\tau$ is the Kohn-Sham kinetic
energy density (KED), expressed in terms of Kohn-Sham orbitals $\phi_i$: 
\be
    \tau = \frac{1}{2} \sum_i^{occup} \left| \nabla \phi_i \right|^2.
    \label{eq:tau}
\ee
The kinetic energy density can in turn be expanded in a gradient expansion 
for slowly varying densities, in terms of $\nabla n$ and $\nabla^2 n$ and
the kinetic energy of the homogenous electron gas:\cite{BJChu,Svendsen}
\be
      \tau = \tau_0 [1 + (5/27)s^2 + (20/9)q]
     \label{eq:tauGEA}
\ee
with $\tau_0 \!=\! \frac{3}{10} (k_F)^2 n$ the Thomas-Fermi approximation to the KED,
i.e., the KED of the locally defined homogeneous electron gas.

This expansion is used to replace $q$ with $\tau$ in the metaGGA family
of functionals.
However, one can also combine Eqs.~(\ref{eq:expandnx}) and~(\ref{eq:tauGEA}) to 
define a gradient expansion of the
exchange hole in the limit of a slowly varying system, with the lowest
order gradient correction of
\be
   \Delta \langle n_{X,\sigma}(u)\rangle = \frac{1}{12}\left[
              \frac{\nabla^2 n}{3} + \frac{4 \left| \nabla n\right|^2}{9n}
          \right]
          u^2,
   \label{eq:nxGEA}
\ee
arguably the more consistent application of the gradient expansion limit.
This correction is consistent with a hybrid of $s^2$ and $q$ with the
choice of $\alpha \!=\! 0.2$.  
Notably, $\alpha$ is rather small, and does not 
give a large role for the Laplacian relative to the gradient, except
in situations where the gradient is negligible.
However, these include important cases such as covalent
bonds, and especially the cusp in the electron density at the nucleus, 
where the Laplacian proves necessary to capture the essential physics.

\subsection{Gauge-invariance of constraints and the local Lieb-Oxford bound}
   GGA's such as the PBE work through imposition of well-chosen constraints --     criteria of reasonability that encourage the construction of forms
   that are robust 
   over a diverse range of systems. 
   For our concept to work, all constraints need to 
   be ``gauge" invariant -- any constraint on GGA must have the same 
   effect for every choice of $\alpha$.  
   If not, the constraint is poorly defined -- the physical information 
    imposed by the constraint in its original gauge (presumably $\alpha\!=\!0$) 
   will not carry over to other choices of gauge.
   Fortunately, most commonly-used constraints
   are energy density gauge-invariant by construction. 
   By the definition of our gauge, 
   the gradient expansion of the functional in the limit of 
   slowly varying density is gauge invariant, 
   and thus also the recovery of the HEG limit in the case of a uniform
   system.  In addition, the 
   behavior of exchange and correlation under uniform scaling of 
   coordinates
   is preserved as we are simply replacing one scale-invariant inhomogeneity
   parameter $s^2$ by another $x$. 
   A final scaling behavior,
   the spin-scaling of the exchange energy, does not involve gradients
   of the density and thus is also unaffected. 

The final class of constraint used in GGA's, the behavior of the 
functional in the limit of large inhomogeneity, is unfortunately not
gauge invariant.  
Unlike scaling laws, these are formulated normally
as constraints on the energy density, not on the energy, 
and the result of imposing such constraints will depend on the choice
of energy density taken.~\cite{footnote1}
More specifically, they are not framed in terms of \textit{systems} that
are highly inhomogeneous, rather than as \textit{regions} of systems, and typically
low-density evanescent regions that have little overall impact on the total
energy. 
There are naturally two such limits, one for exchange and one for correlation,
and we shall be concerned here primarily for exchange.  

The customary constraint on exchange in the limit of high inhomogeneity
($s^2 \!\rightarrow\! \infty$) is the 
Lieb-Oxford (LO) bound,~\cite{Lieb} 
which places a lower bound on the exchange and exchange-correlation
energies for any given density:
\be
     E_{xc}[n]  \ge \lambda_{LO}  E_x^{LDA}[n], 
    \label{eq:LiebOxford}
\ee
with $\lambda_{LO} \!=\! 2.275$.
This is implemented in the PBE and its many offshoots by imposing 
a \textit{local} bound\cite{PBE} on 
the exchange~\cite{LOFootnote} energy density [Eq.~(\ref{eq:exc})] at every point in space:
\be
     e_x(\bfr) = F_X[s^2(\bfr)] e_x^{LDA}[r_s(\bfr)] \ge (1 + \kappa)  
                               e_x^{LDA}[r_s(\bfr)].
     \label{eq:localLiebOxford}
\ee
A choice of $\kappa \!=\! 0.881$ guarantees the overall global bound for 
any density imaginable.

The use of a specific ``gauge" for the energy density in the 
implementation of the LO bound prevents it from being applied to the current 
scheme.  
The local bound is ideal for gradient-only case:
$s^2$ is a positive definite quantity, and is employed to make 
an enhancement factor $F_X$ that is everywhere greater than one, and 
everywhere lowers the energy density.  The
local bound is defined so that the maximum possible lowering of the 
energy density at every point in space will produce an integrated
energy that just hits the global bound, ensuring that it is never 
reached in practice.
The Laplacian however is both positive and negative -- and produces
an $F_X$ that both raises the LDA energy density, typically
inside the atom core, and lowers it, in the asymptotic region outside the atom.
In the latter case, the local LO bound severely limits the possible drop in 
energy, resulting in a net rise in the exchange energy as $\alpha$ is turned
on.~\cite{CancioExlapl} 
In order to maintain the same total energy as the gradient-only case
for all $\alpha$, i.e., to maintain the same \textit{global} bound, 
one needs an $\alpha$-dependent \textit{local} bound that obeys
the local Lieb-Oxford bound only in the $\alpha\!=\!0$ limit.

At the same time, a new perspective on the Lieb-Oxford bound has been given
by Odashima and Capelle which calls into question the relevance even 
of the global bound.
    They show~\cite{OdashimaCapelle} 
     that in practice the global LO bound is never
     approached by any known system (whether atoms or idealized systems
     like Hooke's atom)  -- the practical bound is
     $\lambda \!=\! 1.2$, half the LO bound, seen for He and H$^-$.
     This makes sense -- the LO bound
     only comes into play in regions of extreme inhomogeneity, for
     atomic systems, only in the classically forbidden region
     far from any atom.  Such regions have very low density and very little
     net contribution to the energy overall and so it is next to impossible
     to reach the global LO bound in practice (it is hard to imagine a practical
     system with no classically allowed regions).  Thus in practice no GGA's, 
     even those that break the local LO bound dramatically in the 
     asymptotic limit like the BLYP,~\cite{BeckeGGA,LYP} break the global
     bound for any normal electronic system.  All GGA's get
     He nearly right, that is, the empirical Odashima-Capelle bound 
     $\lambda_{OC} \leq 1.2$, with a practical limit on the 
     XC energy of one-half of the LO bound.  

       
     An additional perspective for the genesis of the GGA approximation
comes from the extended Thomas-Fermi theory of the atom.\cite{PCSB-PRL97,SchwingerPRA1827}  
Basic Thomas-Fermi theory
has a well-known solution~\cite{Scott,March} that becomes exact in the 
limit of large nuclear charge $Z$.  
The solution strictly holds true in the interior
of the atom (where the density is described in terms
of a large number of highly-oscillatory orbitals), failing 
in the classically forbidden region outside the atom, and at the nuclear 
cusp.   
The LDA energy (an extension to the Thomas-Fermi result) asymptotically 
approaches the true atomic energy
as $Z\!\rightarrow\!\infty$, and most of the LDA error can be removed
by an appropriate gradient expansion correction.~\cite{ElliottBurke} 
Thus the large-$Z$ atom can be thought of as a canonical system of 
slowly-varying density, perhaps even the proper reference system to take for 
this limit, as has been done in a recent GGA.~\cite{APBE}

A unifying point of view is that
the low-inhomogeneity limit of the GGA,
given by extended Thomas-Fermi theory, 
describes the interior of the atom, 
whilewhile  extreme inhomogeneity, as measured by
$s^2$ and $q$ exploding to infinity, is the characteristic of its ``surface." 
This is the classically forbidden asymptotic region far from the atom center,
and the cusp in the electron density at the nucleus.  Then the gradient 
expansion limit is the large-$Z$ solution 
since the larger the $Z$, the smaller the surface to volume ratio.  Likewise
the large inhomogeneity limit is the \textit{small}-$Z$ 
limit, where the surface to volume ratio is largest.  
This picture is confirmed by Odashima and Capelle's bound --
the highest value of $\lambda \!=\! E_{xc}/E_{x}^{LDA}$ is 
found precisely for the H$^-$ atom and He,
the two systems that essentially are all surface, and
the value of $\lambda$ for other atoms decreases with $Z$ 
monotonically. 

Thus we suggest ``getting He right", the instinct of empirical GGA's,
is in fact the correct physical constraint needed to determine the 
response of the GGA in the limit of high inhomogeneity. 
We fix $\kappa$ in Eq.~(\ref{eq:localLiebOxford}) as a function
of gauge parameter $\alpha$ by fitting the exchange energy of low-$Z$ atoms,
while setting $\mu$ so as to get the large $Z$ limit. 
This policy neatly places the GGA in its proper context. 
Just as the LDA is fit to a set of numerically exactly solved homogeneous 
many-body systems, we fit the GGA to a set of numerically exactly solved
\textit{inhomogeneous} systems, the neutral closed shell atoms.  
In doing so, we are replacing mathematically motivated constraints
with physical ones, and a bound by precisely known energies.

    \subsection{Ambiguous constraints}
     The discussion above, though satisfactory in
     some respects, obscures a difficulty --
     the value of $\mu$ calculated from perturbation theory for 
     a slowly-varying perturbation of the homogeneous electron gas
     is $\mu\!=\!10/81$, but the value obtained from the extended Thomas-Fermi
     theory of atoms is $\mu \!=\! 0.2604$, over two times bigger.
     Which should be used is not clear, especially for solids, which are 
     composed of atoms and whose conduction bands can 
     approximate a homogeneous system.

     Worse still, the gradient expansion parameter for the 
     overall exchange-correlation energy $\mu \!-\! \mu_c$ is also poorly defined. 
     Quantum Monte Carlo data for the static linear response of the HEG 
     are consistent with the hypothesis that $\mu\!-\!\mu_c$ is exactly 
     zero,~\cite{Moroni,Ortiz} and this point of view
     is taken by many GGA's, particularly the PBE.  
     But perturbation theory about the homogeneous electron gas 
     separately for exchange and correlation~\cite{MaBrueckner} 
     lead to the value $\mu \!-\! \mu_c \!=\! -0.087$.  
     Recent QMC calculations of exchange-correlation holes in real systems suggest that
     $\mu\!-\!\mu_c\!>\!0$.~\cite{CancioLapl,Cancio2RowPaper}
     Other work suggests that neither $\mu$ or $\mu_c$ alone are well-defined
     in the HEG limit, but rather only the combination 
     $\mu\!-\!\mu_c$,\cite{Mattsson1,Mattsson2} a point of view that has some backing
     from XC hole studies of the Si crystal.~\cite{Cancio}
     One can perhaps consider that the extended Thomas-Fermi theory of atoms
     provides a way out of the morass by defining an unambiguous gradient
     correction for exchange, but it has its own problems.
     The extended Thomas-Fermi correction for correlation is not known, nor is
     that for the fourth-order gradient expansion, used in meta-GGA's. 
     
     In practice, there is now an empirical divide between GGA's that
     choose weaker overall corrections, particularly the value of
     $\mu\!=\!10/81$ from perturbation theory and those with stronger 
       corrections
     $\mu\!>\!0.2$ which more closely match 
     the energetics of single atoms.  The former have
     proven to be optimal for reproducing electronic structure, giving 
     the better overall predictions of solid lattice constants and bulk
     moduli than other semilocal DFT's.  The latter do less well for
     structure but perform the best for energetics -- cohesive energies
     and binding energies.  We will focus on two models -- the SOGGA~\cite{SOGGA} which standardizes on a strict adherence to perturbation theory results
for the gradient expansion and exemplifies the weaker gradient-correction model, and
the APBE~\cite{APBE} which standardizes explicitly on the gradient
expansion of the neutral large-$Z$ atom and exemplifies the stronger.

\subsection{Constraints on the potential and nuclear cusp\label{sec:potential}}

So far, we have considered constraints that reproduce those of conventional 
GGA's.  
However, the addition of the Laplacian of the density into our functional 
gains us the ability to do more.  
The Laplacian,
and the related scaleless parameter $q$ has double the range of the
equivalent GGA parameter $s^2$, with the 
possibility of tending to $-\infty$ as well as $+\infty$.  
This allows, and in fact requires, one to fit more constraints to the 
lowest post-LDA functional than would be 
possible by using the gradient alone. 

These considerations come into play particularly 
in the vicinity of the nucleus.  Here
the electron density has a cusp~\cite{Kato} because of the need for an 
infinite local kinetic energy to cancel the $1/r$ singularity in the potential 
energy. 
In this limit, the gradient parameter $s^2$ tends to a finite, small value, with
a nearly universal value of 0.18 for systems with a filled 1s shell.~\cite{TPSS}
In contrast, $q$ tends here to $-\infty$, properly noticing 
a region of extreme inhomogeneity which $s^2$ indicates is slowly varying.
It seems natural then that GGA's have difficulties describing this region -- 
in fact, any functional exclusively using the local 
density and its gradient has an exchange-correlation potential with a 
spurious $1/r$ singularity at the nucleus.   
Since, as far as we know, this cusp is the only physical regime encountered
in ordinary electronic matter in which $q \!\rightarrow\! -\infty$, 
we are free to use the large negative $q$ limit to set constraints on the 
exchange potential at the nucleus without fear of ``contaminating" some other 
feature of density functional space.  
     
To do so, we focus on why the GGA leads to a poor potential near the nucleus. 
The second term in the generating equation for the potential 
[Eq.~(\ref{eq:vxc})] contributes a term to the potential of the form
     \be
       \nabla \cdot \left[\bfs \:\left(\partial F_X / \partial s^2\right) \epsx^{LDA} \right]
     \ee
where $\bfs \!=\! \gradn/2k_Fn$. 
The divergence of the gradient
term in $\bfs$ has a $1/r$ singularity at the nucleus.
To cure this, it is enough to consider an enhancement factor with
an asymptotic expansion of the 
form:\cite{UmrigarGonze}
     \be
     \lim_{q\rightarrow-\infty} F_X(s^2,q) = a + \frac{1}{q}b(s^2) + O\left(\frac{1}{q^2}\right)
      \label{eq:fxcusp}
     \ee
where $a$ is a constant.  The $1/q$ term tends to $r$ at the nucleus where
$q$ tends to $1/r$, guaranteeing a finite potential for any form of 
enhancement factor $b(s^2)$ 
that produces a finite energy density at the nucleus.~\cite{CuspFootnote}

Eq.~(\ref{eq:fxcusp}) gives us in principle the flexibility to satisfy a number
of known constraints on the potential and energy density unambiguously
associated with this limit. 
If we use the conventional definition for the exchange energy-per-particle, 
$\epsilon_X$, as the energy associated with the exchange hole, it is easy
to show that it must be finite at $r\!=\!0$ and have zero slope as well.  
(The LDA and GGA values
for $\epsilon_X(0)$ are naturally finite, but have a finite slope at the nucleus.) 
Known constraints on the exchange potential at the nucleus are that it is 
finite and that it also has zero slope.~\cite{Almbladh}  

A serious consideration for a functional that uses $\lapln$ is that the 
corresponding potential [Eq.~(\ref{eq:vxc})] has
a term with a Laplacian operator, caused by the variation of $E_{xc}$ 
with $\lapln$.  This is much more sensitive 
to changes in the energy density than the 
divergence operator generated by varying $\gradn$.  
The result is a tendency for smooth and apparently reasonable energy 
densities to generate potentials with unphysical and sometimes extreme 
oscillations.~\cite{JemmerKnowles}  
To fix this problem we implement a curvature minimization 
procedure.~\cite{CancioExlapl} 
We construct a DFT model that has a suitable set of 
variational parameters in addition to those fixed by DFT constraints, and 
minimize the curvature integral
\be
     I = \int d^3r \left| \nabla \left(\dexcdlapln\right) \right|^2
\label{eq:curvature}
\ee
numerically for some suitable test density.  
The solution gives the functional form 
of $\partial\exc /\partial\lapln$ with the small possible deviation from a solution of 
Laplace's equation $\nabla^2 (\dexcdlapln) \!=\! 0$
for the set of parameters chosen, and the given test density.  If these are 
chosen well, then the problem of unwanted oscillations can be eliminated 
entirely, as the term that generates them can be made arbitrarily small. 

\subsection{Topology of atoms vis-a-vis the gradient expansion}
\begin{figure}
\includegraphics[clip=true,width=0.40\textwidth]{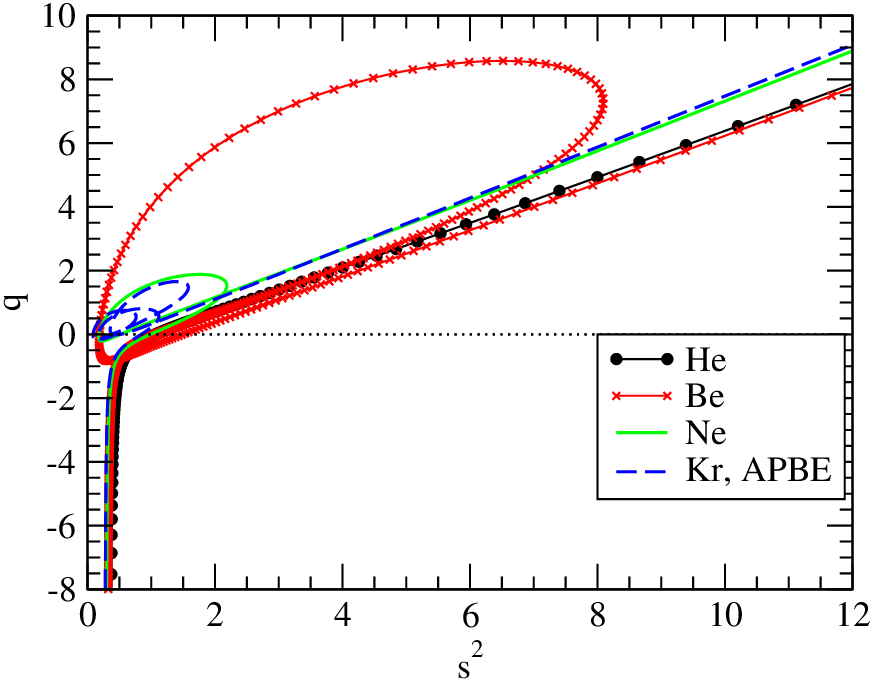}
\caption{\label{fig:s2qparam}
(color online)
The parameter space of the combined gradient expansion parameters $s^2$ and $q$.
Shown are the values of $s^2$ and $q$ typically accessed by physical systems -- represented
here by the He, Be and Ne atoms.  The curves show parametric plots of $s^2(r)$ versus $q(r)$ 
each a function of radial position $r$.  Black with circles is He, red with
crosses is Be, green is Ne, blue dashed is Kr.
}
\end{figure}

Some major issues of this paper can be illustrated by a map of the GEA 
parameter space occupied by the systems we consider, 
specifically the filled-shell atoms He, Be, Ne, and Kr.
Shown in Fig.~\ref{fig:s2qparam} is the value at each point $r$ of a standard 
logarithmic grid of the scale-invariant GEA variable $q(r)$ versus the variable $s^2(r)$ for each atom.  Dots are shown for 
He (black with crosses) and Be (red with circles) in order to show how the 
radial logarithmic grid maps onto the parameter space.  The shortest distances 
from the atom nucleus correspond to the lower left-hand corner of the plot and 
as curves proceed to larger radial distances, they migrate to the 
upper-right-hand corner.
The case of He shows the two limits of behavior that we use to impose 
constraints on our models. The first is a shallow positive slope 
for most of the atom, including the valence peak where $q\!<\!0$, extending out 
to large distances where $s^2$ and $q$ both go to infinity.  Each curve
asymptotically approaches the $q\!=\!s^2$ line in the large $r$ limit, 
but extremely slowly since both variables grow exponentially with $r$; 
the approximately linear behavior holds true anywhere outside the classically
accessible region of the atom.
The second behavior occurs near the nucleus, where $q$ tends to $-\infty$ but 
$s^2$ to a constant.
The resulting curve has an infinite slope.

The topology of the ``phase diagram" of Be is nearly identical to 
that of He, except for a large loop through the positive 
$s^2$ and $q$ quadrant that coincides with the inter-shell region between core 
and valence shells.  
For Ne and Kr, the trend remains essentially
the same.  Not surprisingly, every transition between shells leads to an extra 
loop, three for Kr and one for Ne.  As predicted from the Thomas-Fermi theory 
of large-$Z$ atoms, the larger the nuclear charge, the closer the atom 
approaches globally the slowly varying limit of $q$ and $s^2\!\ll\!1$, and we thus
see the loops get progressively tighter around the origin as the overall 
$Z$ increases.  The behavior near the cusp is essentially universal, as is 
the asymptotic limit far outside the atom, although the latter is approached
at a slightly different rate than for the 
lighter atoms.  
It is interesting to note that the loops are bounded at 
small $q$ (associated with shell peaks) by the asymptotic straight line limit of each atom, while the final $1s$ shell
peak for every atom lines up with the asymptotic limit of the He atom.  

The upshot for DFT of this topological analysis is that we have two
distinct regions, roughly universal in character for atoms, that may
be characterized by the limits of our gradient expansion parameter:
$x\!\rightarrow\!\infty$ and $x\!\rightarrow\!-\infty$.  
This is as long as $\alpha\!>\!0$: at least
some information on $\lapln$ must be included in $x$
in order to detect the latter limit.
At the same time, the quantum oscillations associated with inter-shell
transitions cause a loop behavior that cannot be mapped to a function
of a single variable $x(s^2,q)$, and so any energy density constructed
from such a variable is inherently inexact.  But this behavior could 
potentially be mapped to a general function of $s^2$ and $q$.  As this 
function need be able to handle only a small region near 
$s^2\!=\!q\!=\!0$ for larger 
atoms, it might be handled adequately by the fourth-order gradient expansion, 
which depends explicitly on $s^2$ and $q$.~\cite{Svendsen}

\section{Model and Method \label{sec:methods}}
\subsection{Model of exchange functional}

To develop an exchange functional based upon a hybridization of 
gradient and Laplacian variables,
we start with the GGA.  Taking the form for PBE exchange,
and replacing the GGA inhomogeneity
parameter $s^2$ with the more general $3x(s^2,q)$ of Eq.~(\ref{eq:x}),
we get the exchange enhancement factor
\be
    \Fx[3x(\alpha)] = 1 + \frac{3\mu x(\alpha)}{1 + 3\mu x /\kappa(\alpha)}.
\ee
The model is parameterized through
the parameters $\mu$ and $\kappa(\alpha)$,
(the latter, not being naturally ``gauge-invariant", necessarily dependent 
upon the hybridization factor $\alpha$) and giving rise to the following 
limit cases:
\bea
   \Fx(x) & \rightarrow 1 + 3\mu x,  & \: x \rightarrow 0 \\
   \Fx(x) & \rightarrow 1 + \kappa(\alpha), & \: x \rightarrow \pm\infty.
\eea

A moment's notice shows that this form for $F_X$ will not do, since it has a 
pole at 
finite negative $x$. Because of the electron density cusp,  
$x$ varies from $0$ to  $-\infty$ in the vicinity of the nucleus, guaranteeing that the pole
must be encountered.
However, simple fixes 
such as $\Fx(x) \!=\! 1 \!+\! 3\mu x /\sqrt{1 \!+\! (3\mu x / \kappa)^2}$,  
also fail in this region.~\cite{CancioExlapl}
The 
issue seems to be the infinite variation in the inhomogeneity parameter $x$ 
in a minute region of physical space, $r\!<\!a_0/Z$.
This extreme variation 
gives rise to problems in the X potential when 
the divergence and Laplacian operators of Eq.~(\ref{eq:vxc}) are taken. 

To circumvent this issue, we introduce a ``renormalized" or regulated 
inhomogeneity parameter $\barx$:
\be
     \bar{x} = x \left[1 - \exp{\left(\frac{C\kappa}{3\mu x}\right)}(x<0) \right]
   \label{eq:barx}
\ee
Here $\bar{x}$ has a minimum value of $-C\kappa / 3\mu$ as 
$x\! \rightarrow\! -\infty$, 
and $C$ can then be chosen so that 
the denominator of the functional does not have a pole for negative $x$.
For $x\!>\!0$, we do not need such a regulation -- although the GEA parameter 
tends to $+\infty$ for large radii, it does so over an infinite range
and seems to lack the serious instabilities encountered near the cusp.

In the end, we use with the following 
modified functional form for exchange:
\be
    F_X(x) = 1 + \frac{ 3\bar{\mu}(x) \bar{x} }
                      { \sqrt{1 + \eta (3\mu \bar{x}/\kappa) + \left(3\mu \bar{x}/\kappa\right)^2} }
    \label{eq:fxmodGGA}
\ee
where $\barx$ is the regulated version of $x$ described by Eq.~(\ref{eq:barx}).
The renormalization of $\mu$ in the numerator,
\be
     \bar{\mu} = \mu \left[1 - A \exp{(C\kappa / 3\mu x)}(x<0) \right],
    \label{eq:barmu}
\ee
gives an additional degree of control allowing us to fit a further constraint 
on the potential at the nucleus.
In addition to conventional DFT constraint parameters 
$\mu$ and $\kappa$, the model uses
variational parameters $C$, $A$ and $\eta$ which can be 
chosen to optimize curvature in the exchange potential.
This form has the same behavior as the PBE for $x\!\rightarrow\!0$ and
$x\!\rightarrow\!\infty$ limits,
and in the $x\!\rightarrow\!-\infty$ limit reduces to
\be
     1 - \frac{(1-A)\kappa C}{\sqrt{1 -\eta C + C^2}} + O(1/x)
     \label{eq:fxlimit}
\ee
which has the desired form [Eq.~(\ref{eq:fxcusp})] as long as the 
denominator in the second term is nonzero and real.

\subsection{Calculation of energy and potential}

As a starting point of our functional development,
we focus on functionals that exactly reproduce the limiting behaviors
of known GGA's.  
The APBE~\cite{APBE} and SOGGA~\cite{SOGGA} variants of the PBE are 
chosen as representative of two extremes of interpretation of 
conflicting information about DFT constraints.
We keep the $\mu$ value of each model and determine values for 
$\kappa(\alpha)$ by requiring 
that the new functional reproduce for any $\alpha$ the exchange energy of the
original for the
Neon atom, representing the small-$Z$, and thus high-inhomogeneity limit. 
This strategy gives
rise to two new models, a modAPBE and a modSOGGA, defined for the hybrid 
inhomogeneity parameter $x$ with mixing coefficient of $\alpha\!=\!0.2$.  
A third, SOGGA-q, represents our best attempt at producing a Laplacian-only model. 
The parameters used to define these models and their GGA equivalents 
are shown in Table~\ref{table:models}.

\begin{table}[ht] 
\begin{centering}
\begin{tabular}{|l|l|l|l|l|l|l|}
\hline
Model & $\alpha$ & $\mu$  & $\kappa$ & $\eta$ & $C$ & $A$ \\
\hline\hline
 LDA       &         & 0.0     &       &      &      &\\
 SOGGA     &  0.0    & $10/81$ & 0.552 & 2.0  &      & \\
 APBE      &  0.0    & 0.26037 & 0.804 & 2.0  &      & \\
 SOGGA-q   &  1.0    & $10/81$ & 0.552 & 2.0  & 1.00 & 1.00  \\
 ModSOGGA  & 0.2 & $10/81$ & 1.104 & 3.0  & 0.117 & 0.129 \\
 ModAPBE   & 0.2 & 0.26037 & 1.55  & 3.0  & 0.252 & 0.366 \\
\hline
\end{tabular}
\caption{\label{table:models}
Parameters used in defining the DFT models discussed in this paper,
as defined by 
Eqs.~\ref{eq:x}, \ref{eq:barx}, \ref{eq:fxmodGGA}, and \ref{eq:barmu}.
}
\end{centering}
\end{table}

The remaining variables $\eta, C, A$ 
are used to constrain the potential, $\eta$ in the region outside
the nuclear cusp ($r\!>\!a_0/Z$), $C$ and $A$ inside.  Minimization of the 
curvature [Eq.~(\ref{eq:curvature})] was the primary constraint with 
additional considerations from the known behavior of the potential 
at the nucleus.
The curvature integral $I$ must in practice
be optimized for some test density or densities and for this purpose we
use that of the Be atom.
The features that we find produce unphysical curvature in the exchange 
potential of atoms, the nuclear cusp and the 
exponential tail in the classically forbidden region outside the atom, 
are basically universal, whereas Be in addition suffers from
high levels of inhomogeneity in the core-valence transition that are
an additional source of trouble. 
Optimization of the potential for this case
produces nearly optimal results for the other atomic systems we have studied.  

As reference data for small atoms, 
we use exact Kohn-Sham densities and exchange potentials
\cite{UmrigarGonze,Filippi}
which are evaluated on a standard logarthmic grid.
Exchange energies are also evaluated for larger closed-shell atoms using
the APE pseudopotential generator~\citep{APE} in all-electron mode and 
the APBE exchange-correlation functional. 
Laplacians and gradients that appear in the exchange potential are evaluated
numerically by the method of Bird and White,~\cite{WhiteBird} evaluated
on the grid with Lagrange interpolating polynomials.  To check 
numerical calculations, a few were made using Slater-type orbitals
for which analytic values for derivatives could be obtained.  
The errors in numerical derivatives were negligible (relative errors of order 
$10^{-9}$ for polynomials of order 11 or 13 and standard grids.)

\section{Results\label{sec:results}} 

\subsection{Optimization of functional parameters}

Our first results are the values of the parameters, shown in Table~\ref{table:models},
that meet the constraint conditions on energy and potential --
matching known GGA's for large-$Z$ atoms (representative of 
slowly-varying densities), small-$Z$ atoms (quickly-varying
densities), minimizing the curvature
integral $I$ to produce a physically reasonable potential at all $r$, and
producing a finite exchange potential with zero slope at the nucleus.

Being a gauge-invariant parameter, $\mu$ 
takes the value of the 
corresponding GGA whose constraints we are trying to duplicate, $10/81$ for
the SOGGA and $0.2604$ for the APBE, independent of the choice of $\alpha$ taken.  
To satisfy the other constraints proved tricky given the interconnection between
parameters within the form given by Eq.~(\ref{eq:fxmodGGA}).
The exchange energy depends primarily on the region outside the nucleus and can
be decoupled from the parameters $C$ and $A$ that characterize large negative $x$.
For any fixed gauge choice $\alpha$, the requirement that our functional
reproduce the exact exchange energy of the Ne atom, our large-inhomogeneity constraint,
defines a continuum of possible values $\kappa(\eta)$ for the two remaining free
parameters. 
The exchange potential develops unphysical oscillations at the outside edge
of the valence shell for $\eta \!<\! 2$, which are particularly large for the Be atom.
At the same time larger values of $\eta$ lead to problems at the nucleus,
moving the limiting value of $F_X$ [Eq.~(\ref{eq:fxlimit})] closer to a singularity.  
There seems to be a conflict in which optimizing the potential
for the one region of space comes at the cost of unphysical curvature in the other.  
A compromise that works for the natural gauge $\alpha\!=\!0.2$ taken
from gradient expansion of the exchange hole is to take $\eta \!=\! 3$ and 
$\kappa$ twice the value derived from the local LO bound.  
This almost perfectly reproduces GGA exchange 
energies of filled-shell atoms for our modification of both the SOGGA and the APBE
GGA's, while having essentially the same curvature in the Be potential as the 
optimal case $\eta \!\sim\! 5$.  
This is somewhat of a surprise since $\eta$ was 
designed to help prevent a pole in $F_X$ at the nucleus, while it functions
best to remove curvature in the valence shell.
Finally, for $\alpha\!=\!1$, the Laplacian-only theory, it was difficult to find a 
value of $\kappa$ for which one could match the exchange energy of the 
GGA and retain a reasonable potential; we had to settle for a reasonably smooth
potential and an exchange energy that was not much better than the LDA.

The parameters $C$ and $A$ play a leading role for large negative 
$x$, roughly $r<0.1~a_B/Z$, and give us variational freedom to minimize curvature 
in this region and impose an additional constraint on the
X potential at the nucleus -- for example the zero slope of the potential 
observed in numerical data.~\cite{Filippi,Almbladh} 
However the minimum curvature in the potential in this region is achieved by introducing
a pole in the energy density and potential at the nucleus -- the constant $a$
in Eq.~(\ref{eq:fxcusp}) tends to infinity as the curvature is minimized.
We retain the singularity of the GGA potential, while adding
a new singularity in the energy density!
The strength of the pole 
decreases if $\eta$ is reduced, putting
curvature back in the large $r$ limit, but it occurs at any value of 
$\eta$, $\mu$ or $\kappa$ other than
zero.  Nevertheless, as shown below, this pole can be avoided by 
accepting a modicum of unphysical curvature in the potential at finite radius.
Within the limitations of our current form for $F_X$, competition 
between incompatible constraints limits our ability to meet all our goals
for the exchange potential at the nucleus.

\begin{figure}[ht]
\includegraphics[clip=true,width=0.30\textwidth]{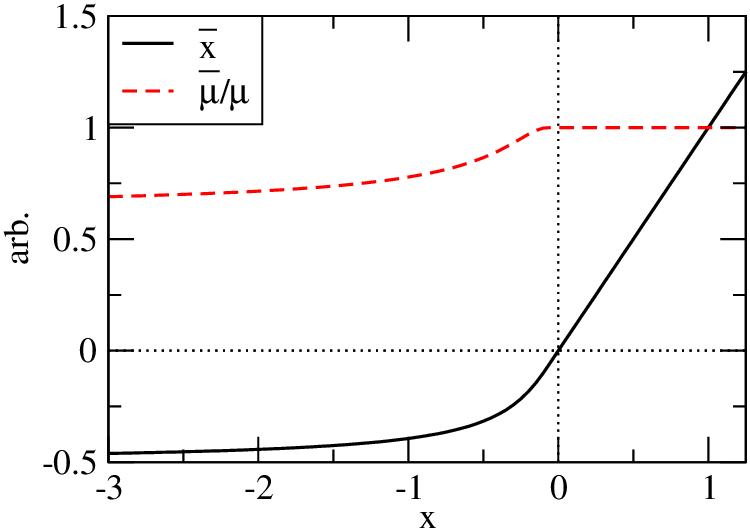}
\caption{\label{fig:xbar}
(color online)
Regularized gradient expansion variable $\barx$ (black) 
and linear coefficient $\bar{\mu}$ divided by $\mu$ (red, dashed), for the 
modAPBE functional.
}
\end{figure}

\begin{figure}[ht]
\includegraphics[clip=true,width=0.35\textwidth]{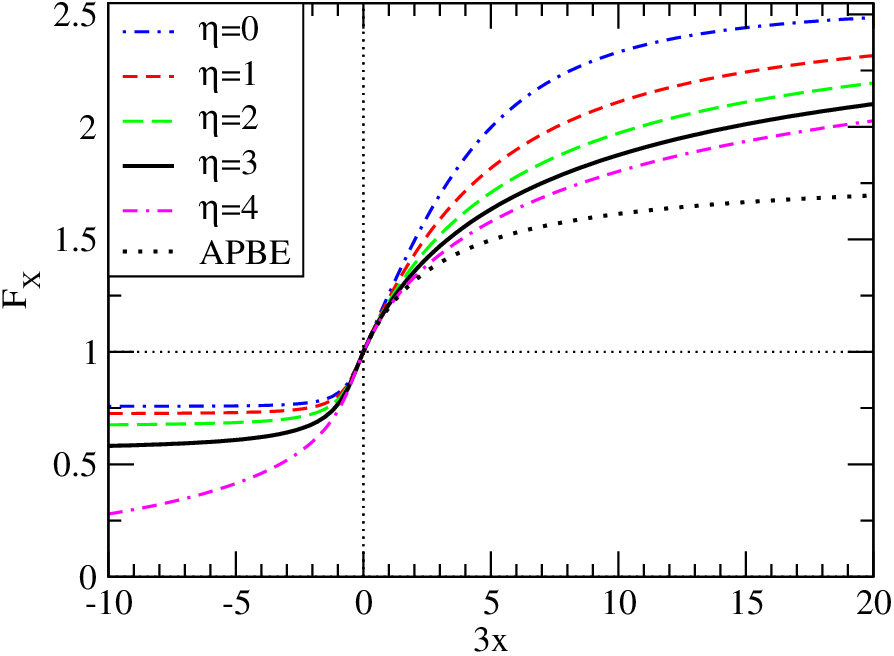}
\caption{\label{fig:fxmaps}
(color online)
The enhancement factor $F_X$ for the modAPBE functional, as a function of
gradient expansion parameter $3x$, for several values of curvature
parameter $\eta$, with the other parameters fixed.  Solid line shows optimal 
value of $\eta$.  Dotted line is $F_X$ for the APBE GGA.
}
\end{figure}

Figs.~\ref{fig:xbar} and~\ref{fig:fxmaps} display some of the features
of the enhancement factor introduced in this paper.
Fig.~\ref{fig:xbar} shows the ``regulated" gradient expansion
variable $\xbar(x)$ and the ratio of ``regulated" to unregulated 
gradient coefficients $\mubar(x)/\mu$, both used 
in $F_X$ [Eq.~(\ref{eq:fxmodGGA})] to optimize its 
form near the nucleus. The former is equal to $x$ for $x\!>\!0$ but saturates
to a finite value for large $x\!<\!0$, thus avoiding a pole in $F_X$.  
The latter shows a renormalization of the gradient coefficient $\mu$ to 
$\mu(1\!-\!A)$ for large $x\!<\!0$. 
Fig.~\ref{fig:fxmaps} shows the resulting enhancement factor $F_X(x)$ 
versus $3x$ for the modAPBE, for several values of the potential 
optimization parameter $\eta$ and 
in comparison, the enhancement factor for the APBE. (Note that $3x$ maps
to $s^2$ if the mixing parameter $\alpha$ is turned off.)
All the forms have the same linear correction $\mu$ and are thus identical
at small $x$.
The maximum possible value of $F_X$ for the modAPBE is $1\! +\! \kappa\! =\! 2.55$, 
larger than that of the APBE, and compensates for the ``dis-enhancement" of
the exchange energy that occurs for negative $x$.
The regulating effects of using $\xbar(x)$ as a variable are seen in the 
more rapid saturation of $F_X$ for negative values of $x$ than for positive,
and notably in the absence of a pole in $F_X$ for any value of $x$.
The enhancement factors with smaller values of $\eta$ have a more kinked
form at large $x$, with a larger first derivative, and thus greater curvature 
$I$.  The smoothest forms
($\eta \!\sim\! 5$ for the modAPBE) cause an unrealistic positive $\epsx$ at
the nucleus, while the chosen value $\eta\! =\! 3$ gives a nearly correct
value for $\epsx(0)$ for He.

\subsection{Energies}
\begin{figure}
\includegraphics[clip=true,width=0.40\textwidth]{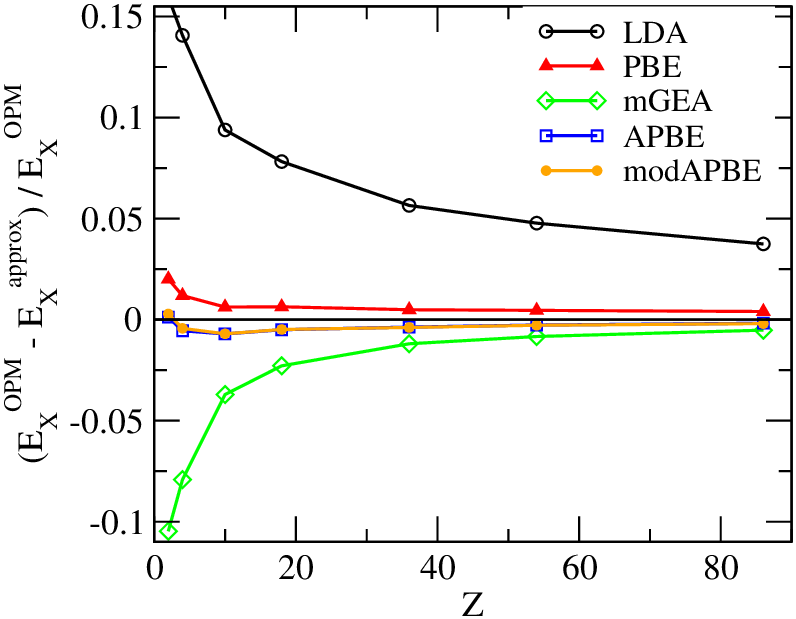}
\caption{\label{fig:largeZscaling}
(color online)
Fractional error in exchange energy of closed shell atoms for various DFT 
approximations, relative to optimized potential method (OPM) 
values,~\cite{EngelVosko1,KPB}
plotted versus nuclear charge $Z$. 
}
\end{figure}

Figure~\ref{fig:largeZscaling} shows the 
relative error of the exchange energy with respect to the essentially
exact optimized potential method (OPM)~\cite{EngelVosko1,KPB} for filled 
shelled atoms from He to Rn.  Shown are the results for the LDA, the 
PBE and APBE forms of the GGA, the modAPBE defined in this paper and finally
the modified gradient expansion approximation model (mGEA) of 
Ref.~\onlinecite{ElliottBurke}. 
All energies are calculated with the non-relativistic Schrodinger equation
so as to highlight the large-$Z$ scaling limit of Thomas-Fermi theory.
The energies for the mGEA and modAPBE functionals are calculated using the APBE density; all others are determined self-consistently.
We estimate the use of non-consistent
densities to have a very small effect.  For example, the PBE exchange energy 
using the OPM-derived charge densities~\cite{EngelVosko1} 
is 0.03-0.05\% off from that using self-consistent densities.

Perhaps the most interesting data here is that of the mGEA, which uses
the gradient expansion enhancement factor
\be
   F_X(s^2) = 1 + \mu s^2
   \label{eq:mGEA}
\ee
with the value of $\mu\! =\! 0.26037$ 
obtained by fitting the gradient expansion to the 
asymptotic large-$Z$ limit of the exchange energy of atoms.  
Within the extended Thomas-Fermi model of the atom, the LDA value
for the exchange energy scales as $Z^{5/3}$ with nuclear charge $Z$, 
while the GEA correction scales as $Z$, and the next higher term as $Z^{2/3}$.
Thus, the relative error in the LDA should decrease as $Z^{-2/3}$ as 
$Z \!\rightarrow\! \infty$ and the relative error in the GEA, even faster, as $Z^{-1}$.
This trend is clearly seen in Fig.~\ref{fig:largeZscaling} -- 
the LDA undershoots the magnitude of the exchange energy for 
He by about 16\%, but declines to 3\% for $Z\!=\!86$.
This error is essentially eliminated for large
$Z$ by the gradient expansion with the mGEA value for $\mu$.  
If one uses the ``traditional" value, $\mu\!=\! 10/81$, determined from perturbation theory
about the HEG, only half of the LDA error gets removed.  

The main error of the mGEA, the poor treatment of the large-$s^2$ corrections 
in the asymptotic region of the atom, primarily affects the smaller $Z$ atoms, 
since this region constitutes a portion of the total charge density -- the tail of the valence shell --  that decreases as $1/Z$ as $Z \!\rightarrow\! \infty$.
The mGEA systematically overshoots the OPM in magnitude, consistent with 
a lack of attention to the Lieb-Oxford bound in the GEA form, with gradient 
corrections unbound from below as 
$r \!\rightarrow\! \infty$, as shown in Fig.~\ref{fig:vxHe}(b).  
The correction of this 
failure at large $s^2$ or small-$Z$ is the key improvement of GGA exchange over 
the gradient expansion.  Both GGA's shown, the PBE and the APBE, 
improve dramatically upon the GEA in the low-$Z$ limit.
The PBE with $\mu\! =\! 0.2195$ 
is not quite as good a fit as the APBE, which has the same Lieb-Oxford 
bound limit $\kappa$, but the larger mGEA value of $\mu$. 
The APBE wins out especially at large $Z$, where it has half the 
error of the PBE.  

The modAPBE introduced here has the mGEA value of $\mu$ and a value of
$\kappa$ (=1.55) adjusted to fit the APBE exchange energy for He and Ne. 
With these constraints, it
duplicates the trend in exchange energy for the APBE for all atoms 
with perhaps a slight degradation in the percent error at high $Z$.  
This suggests that the physics of the exchange energy of closed shell-atoms
is fairly simple and almost completely determined by GGA constraints.  And
once we determine the right set of constraints to use -- ones that are invariant with the
choice of energy density gauge -- the density
functional parameter used to implement them becomes largely irrelevant.

The GEA form as applied to our hybrid parameter $x$ is 
$ F_X(x)\! =\! 1 \!+ \!3 \mu x $
and by construction should give identical exchange energies 
regardless of the choice of $\alpha$ in Eq.~(\ref{eq:barx}),
given the interchangeability of the gradient and Laplacian variables to 
this order in the gradient expansion.  
It should also give identical exchange potentials, 
despite the very different equations used to generate gradient and Laplacian 
contributions to the overall potential [Eq.~(\ref{eq:vxc})].  
We have verified this for the mGEA energy using the $\alpha\!=\!0$ ($s^2$ only) 
and $\alpha\!=\!1$ ($q$ only) gauges and APBE charge densities.  Both energies
and potentials are indistinguishable up to numerical error (7 to 11 
significant figures), providing an excellent check for the numerical
methods used to calculate them.  

Table~\ref{table:energy} 
shows the exchange energy, the value of the exchange potential near $r\!=\!0$, and the 
curvature integral $I$ [Eq.~(\ref{eq:curvature})]
for some of the DFT models we have discussed so far.  
The SOGGA and APBE 
define weak and strong GGA corrections to the LDA respectively,
the modSOGGA and modAPBE are functionals designed to reproduce these GGA's 
for atomic systems, and the SOGGA-q is our best Laplacian-only model.
The calculations are performed for exact Kohn-Sham densities~\cite{UmrigarGonze,Filippi} 
and compared to exact exchange potentials.\cite{Filippi}

\begin{table}[ht] 
\begin{centering}
\begin{tabular}{|l|l|l|l|l|}
\hline
Atom  &  Model & $E_x$ & $V_x(0)$ & $I$ \\
\hline\hline
He  &  LDA       & -0.883   & -1.509 & \\ 
    &  SOGGA     & -0.961   & -14.88 & \\ 
    &  APBE      & -1.02995 & -28.77 & \\ 
    &  SOGGA-q   & -0.9017  & -9.966 & 1.5e-03 \\ 
    &  modSOGGA  & -0.9599  & -2.409 & 1.87e-04 \\ 
    &  modAPBE   & -1.0280  & -5.52  & 4.52e-04 \\ 
    &  KS        & -1.02457 & -1.688 & \\ 
\hline
Be  &  LDA      &  -2.321  & -3.230 & \\ 
    &  SOGGA    &  -2.513  & -98.37 & \\ 
    &  APBE     &  -2.688  & -198.0 & \\ 
    &  SOGGA-q  &  -2.368  & -25. & 1.5e-1 \\ 
    &  modSOGGA &  -2.512  & -5.48 & 2.7e-3 \\ 
    &  modAPBE  &  -2.6844 & -13.2  & 9.4e-3 \\ 
    &  KS       &  -2.674  & -3.126 & \\ 
\hline
Ne  &  LDA     & -11.021  & -8.391 & \\ 
    &  SOGGA   & -11.621  & -238.4 & \\ 
    &  APBE    & -12.209  & -480.3 & \\ 
    &  SOGGA-q & -11.334  & -70.   & 1.5e-3 \\ 
    & modSOGGA & -11.623  & -14.8  & 2.22e-04 \\ 
    & modAPBE  & -12.2079 & -36.5  & 5.38e-04 \\ 
    &  KS      &          & -7.984 & \\ 
    &  HF      & -12.11   &        & \\ 
\hline
\end{tabular}
\caption{\label{table:energy}
Exchange energy, potential at the nucleus, curvature integral for GGA's and Laplacian-based models (in hartrees), 
evaluated using exact Kohn-Sham densities.~\cite{UmrigarGonze,Filippi}
KS are corresponding results from exact Kohn-Sham density functional theory,
and HF from exact Hartree-Fock calculations.\cite{ClementiRoetti} 
}

\end{centering}
\end{table}

The SOGGA-q uses the local LO bound value for $\kappa$, but with an energy
density gauge ($\alpha\!=\!1)$ quite different from that of a GGA ($\alpha\!=\!0$).  
The result is a much smaller exchange energy than the SOGGA, close to the LDA, and
shows the effect of 
uncritically using for one gauge of the energy density a constraint defined with respect 
to another.
For the two modGGA functionals, 
the exchange energy of the respective gradient-only GGA has been matched
by doubling the large-inhomogeneity parameter $\kappa$.  
For the curvature integral $I$, a value $\sim 1\times10^{-4}$ indicates a potential which is a monotonic 
convex function, as expected for the X potential for He.  
The SOGGA-q potential for He is thus not quite
optimal but the modAPBE and modSOGGA ones are stable and 
free of spurious oscillations as indicated by their low $I$.  
Given the definition of $I$, [Eq.~(\ref{eq:curvature})],
it is natural that as the $q$ component of the inhomogeneity variable $x$ 
is turned on, $I$ becomes larger, so that it is consistently at least an order 
magnitude higher for the SOGGA-q model with $\alpha\!=\!1$ as compared to the 
modSOGGA with $\alpha\!=\!0.2$. 
The higher values of $I$ for Be are physically relevant as will be shown below.

The value of the X potential shown in Table~\ref{table:energy} is not exactly for 
$r\!=\!0$, but rather the smallest radius ($\sim 10^{-3}/Z$) in the numerical grid 
used to defined the atomic density.  
The value of $V_X(0)$ in hartrees obtained from the exact Kohn-Sham calculations
varies with $Z$ approximately as $-0.8Z$.  
The GGA's evaluated at this radius are well on their way to $-\infty$ while 
the cusp-corrected modifications of the GGA are clearly finite but lower than
the exact values.
The best case scenario is the modSOGGA for which $V_X(0)$ varies roughly as $-1.4Z$. 
The stronger gradient correction used by the APBE and modAPBE
leads to worse behavior at small $r$: the APBE singularity is more noticeable
than that of the SOGGA and the modAPBE $V_X(0)$ is almost five times deeper 
than the exact value.

\subsection{He atom}

\begin{figure}
\includegraphics[clip=true,width=0.40\textwidth]{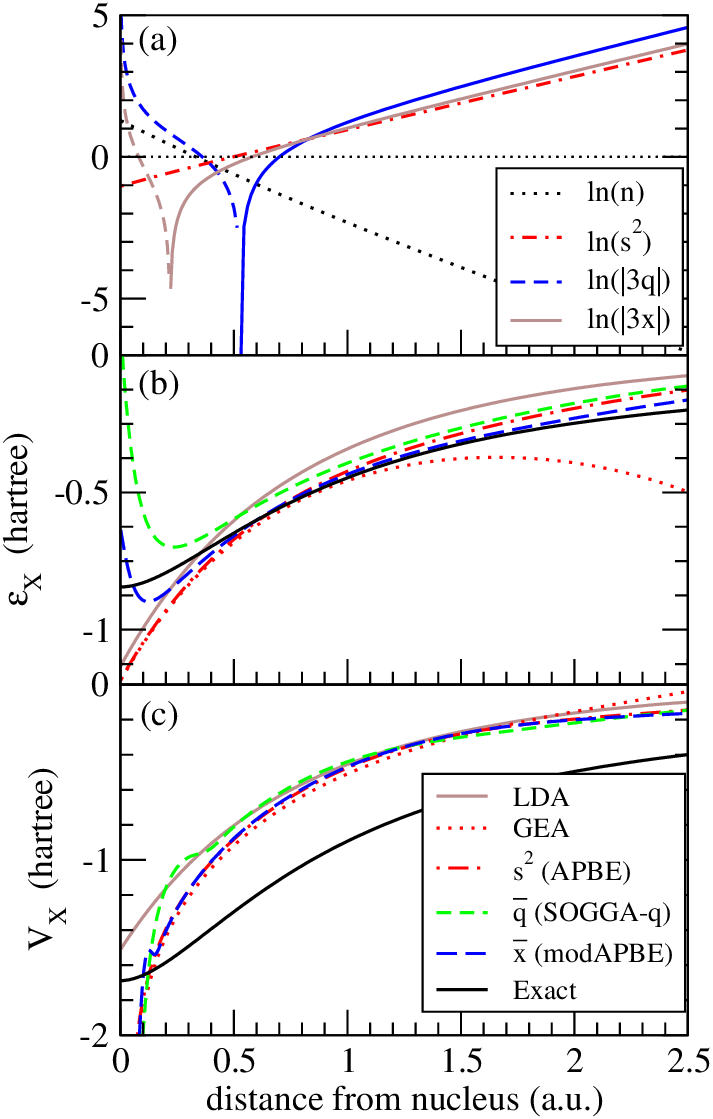}
\caption{\label{fig:vxHe}
(color online)
GEA Density functional parameters (a), exchange energy-per-particle (b), 
exchange potential (c) for the He atom, as function of distance from 
the nucleus.  In (a), the logarithms of the density (dotted), GEA parameter
$s^2$ (dash-dotted) the absolute value of $3q$ (dark solid line for $q\!>\!0$ and dashed
for $q\!<\!0$) and the hybrid parameter $3x$ (light solid and dashed) are 
plotted.  In (b) and (c), quantities are derived from the LDA (light solid), the
mGEA~\cite{ElliottBurke} (dotted) the APBE GGA~\cite{APBE} (dot-dashed), 
two functionals defined in the text -- 
the SOGGA-q using the Laplacian
variable $q$ (dashed), the modAPBE using the variable $x$.
These are compared to numerical values for the exact integral expressions 
for $V_X$ and $\epsx$ (dark solid). 
All quantities are determined using the charge density of Ref.~\onlinecite{UmrigarGonze}.
}
\end{figure}

Fig.~\ref{fig:vxHe}(a) shows logarithms of the density and the 
gradient-expansion parameters $s^2$, $3q$, and $3x$, as
a function of radial distance for the He atom.  The density is 
approximately exponential, leading to a straight line for its logarithm. 
The parameter $s^2$ for an exponential is also exponential and 
its logarithm increases linearly from a value of $0.18$ near the nucleus.  
The variables derived from the 
Laplacian, $3q$ and $3x$, are more complex, since the Laplacian changes sign 
at $r\sim a_0/Z$ and
diverges to $-\infty$ near the nucleus.  

The parameter $s^2$ is defined so that a value appreciably less than one 
(zero on the log plot) indicates a region of slowly-varying density for 
which the GEA presumably should be valid.  
Based on this, one might expect the GEA to be valid for
all positions within a scaled Bohr radius $a_0/Z$ and gradually diverge from the correct result at 
large $r$.
The Laplacian-based equivalent to $s^2$, $3q$, 
diverges both at the nuclear cusp 
and in the asymptotic limit and thus is a more accurate indicator of where 
the GEA fails.  
The hybrid GEA parameter $3x$ represents a compromise between the two 
cases.  It is quite noticeably close to the $s^2$ measure almost everywhere, a 
consequence of the small value of $\alpha$ that comes from the Becke 
exchange-hole analysis.  However, it diverges at the nucleus, so 
that it holds true to the correct physics in this region.

Figs.~\ref{fig:vxHe}(b) and (c) show the exchange
energy-per-particle 
and potential of the He atom as a function of distance from the nucleus, 
using several of the density functional theories discussed in this paper.
These include the LDA, the mGEA,
the APBE, presumably the most 
accurate conventional GGA for atomic systems,
and two models based on the Laplacian of the density:
the SOGGA-q, using the regularized Laplacian factor $3\bar{q}$ 
and the modAPBE, using the regularized hybrid of gradient and 
Laplacian, $3\bar{x}$.
For this system, exact values of $\epsx$ and $V_X$ may be easily obtained 
given that the exchange energy for a two-electron singlet is 
simply the removal of the self-interaction energy of each electron and is 
obtainable from the Hartree potential.  A value of 
$\epsx(\bfr)\!=\!-V_H(\bfr)/4 \!=\!-\int\; n(\bfr')/4\left|\bfr-\bfrp\right| d^3r'$ 
obtains this result.  (This is not a unique definition of $\epsx$
of course, but is directly derived from the exchange hole 
of the spin singlet, and thus the definition we will be interested in.)
Likewise, the potential $V_X$, shown in (c), is equal to $-V_H/2$.  
As before, we use the exact Kohn-Sham density for He for this 
purpose.~\cite{UmrigarGonze}

The LDA is a not unreasonable ballpark result for $\epsx$, but it has
qualitatively wrong behavior (a cusp) at the nucleus and is shifted upwards from the 
$-1/2r$ limiting behavior of the true $\epsx$. The mGEA improves 
upon the LDA at high density; far from the nucleus, where $s^2$ diverges, 
it deviates severely from the correct behavior.  
The conventional GGA, because of its adherence to the 
local Lieb-Oxford bound, corrects this extreme behavior to provide a reasonable
global improvement to the LDA energy-per-particle and energy.  
Of the three GGA-type models, gratifyingly it is the modAPBE, 
the hybrid gradient-Laplacian model
designed to come as close as possible to the energy of 
the XC hole, that most closely adheres to the exact values for $\epsx$.
It does so quite closely for the intermediate distances from the nucleus
which contribute most to the total energy.  
As $r \!\rightarrow\! \infty$, it decays exponentially like
the LDA and GGA, but at a slower rate.

Fig.~\ref{fig:vxHe}(b) reveals why the exchange energy for the SOGGA-q
fails to match that of the SOGGA in Table~\ref{table:energy} and illustrates the need for 
treating $\kappa$ as a function of the energy-density gauge $\alpha$.
DFT models that rely on $\lapln$ have higher energy density than the 
LDA near the nucleus where $\lapln<0$; in contrast GGA's have a lower
energy density.  To match the GGA integrated exchange energy, it is then necessary to 
lower the energy density elsewhere -- specifically in the asymptotic large-$r$ 
region.  This is a region of high inhomogeneity where the LO bound kicks in and
prevents the energy density from being lowered.
Unless it is ignored, the exchange energy of a 
modGGA cannot be made to match that of the corresponding GGA, but must
necessarily be too high.  

In Fig.~\ref{fig:vxHe}(c) are shown the He exchange potentials for the models 
discussed in Fig.~\ref{fig:vxHe}(b).  The exact model is cuspless and
has an asymptotic limit of $-1/r$, and is otherwise largely featureless.
The LDA exchange potential is proportional to $n^{1/3}$ and thus 
has a cusp at $r\!=\!0$ and decays exponentially at large $r$.
The result is the well-known,~\cite{JG} roughly constant shift upwards in 
energy of the LDA with respect to the exact potential.
The various
gradient-correction models adhere closely to the LDA at intermediate distances,
diverge slightly from it at large distance and slope sharply downwards at the
nucleus.  The GEA diverges to positive infinity at very large distances.
The modAPBE, although it produces a very accurate fit to $\epsxc$,
produces a potential that is not noticeably
different from the APBE from which it is derived.

\begin{figure}
\includegraphics[clip=true,width=0.37\textwidth]{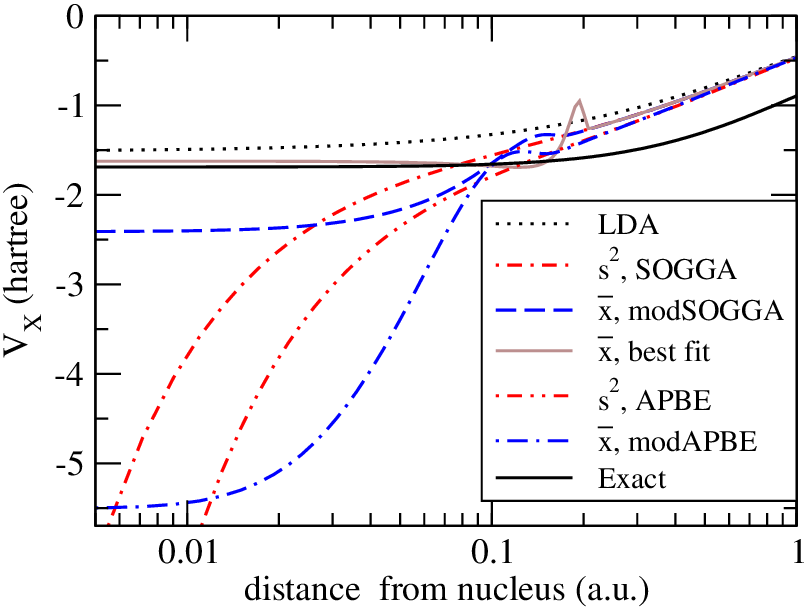}
\caption{\label{fig:vxHecusp}
(color online)
X potentials for the He atom at the nuclear cusp.  Dotted is LDA, exact is black
solid, SOGGA GGA\cite{SOGGA} is double-dot-dashed, APBE GGA~\cite{APBE} is dot-dashed, 
modSOGGA long-dashed, modAPBE is short-dashed.  Finally a version of the 
modSOGGA optimized to reproduce the exact potential at the nucleus is shown as grey 
solid line (brown online).
}
\end{figure}

Fig.~\ref{fig:vxHecusp} shows the behavior of the DFT potentials shown in
Fig.~\ref{fig:vxHe}(c) in the immediate vicinity of the nucleus.
The pole in the GGA potential occurs in the 
``deep cusp" region: $r\!\leq\!0.05a_0$ for He,
and roughly $r\!\leq\!0.10a_0/Z$ in general. 
The SOGGA GGA, a weaker correction to the LDA, is understandably better 
behaved than the more aggressive APBE.  
The corresponding modSOGGA and modAPBE potentials are shown as dashed lines, 
and show the benefit of including the Laplacian of the density.
The modSOGGA is finite at the nucleus with a quite reasonable value
for $V_X(0)$ and is explicitly cusp-free like the true potential.  
The modAPBE, though finite at the nucleus, is much further from the 
true potential, at best marginally better than the corresponding GGA.  
In both cases, the problem of an unphysical potential at the nucleus
does not disappear as much as change form -- the pole created by the GGA spreading out 
spatially into an unphysical dip.  

The consequence of conflicting optimization strategies for the cusp region -- 
either to minimize the curvature or fit the potential at $r\!=\!0$ is 
demonstrated by a model (grey, brown online) in which the cusp correction 
parameters are chosen 
to provide a best fit to the exact X potential in the vicinity of the nucleus.
This very close fit comes at a cost of a large fluctuation in the potential, 
of order a rydberg in energy, in the region where the regularized 
variable $\bar{x}$ approaches zero.  
Conversely, completely minimizing curvature comes at the cost of creating 
a pole at $r\!=\!0$, one which is in fact worse than the corresponding GGA 
pole.  The optimal curve shown in the figure is thus a somewhat arbitrary 
balance between these two competing effects -- minimizing the
magnitude of $V_X(0)$ while keeping the slope in $V_X$ 
nonnegative for $r\!>\!0$.

\subsection{Beryllium and Neon}
\begin{figure}
\includegraphics[clip=true,width=0.40\textwidth]{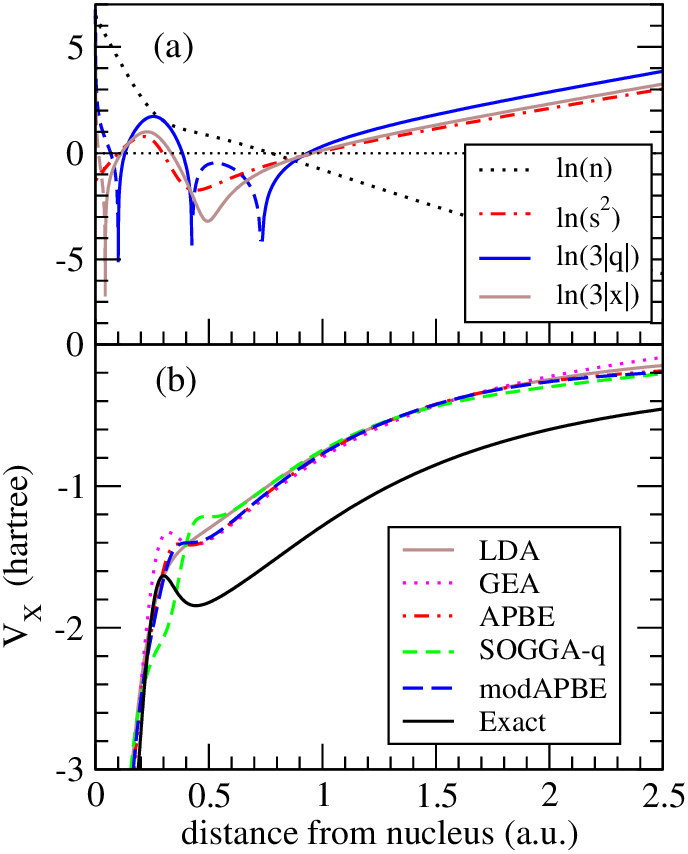}
\caption{\label{fig:vxNe}
(color online)
GEA Density functional parameters (a), and exchange potential (b)
for the Ne atom, as function of distance from 
the nucleus.  Quantities plotted are the same as for Fig.~\ref{fig:vxHe}(a)
and (c).
}
\end{figure}

\begin{figure}
\includegraphics[clip=true,width=0.40\textwidth]{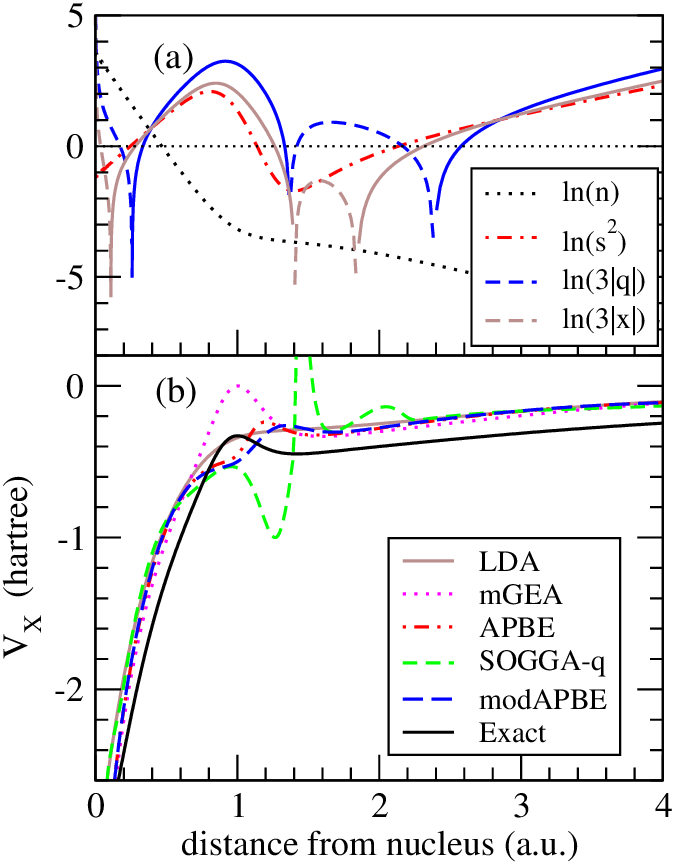}
\caption{\label{fig:vxBe}
(color online)
GEA Density functional parameters (a), and exchange potential (b)
for the Be atom, as function of distance from 
the nucleus.  Quantities plotted are the same as for Fig.~\ref{fig:vxHe}(a)
and (c).
}
\end{figure}

Systems of considerably more interest than He 
are those of Ne. 
shown in 
Fig.~\ref{fig:vxNe} and Be, shown in Fig.~\ref{fig:vxBe}.   
These are the smallest closed-shell atoms 
that contain the topological feature of a transition between two shells. 
As shown in Fig.~\ref{fig:s2qparam}, the sequence He, Ne, Be represents
a progressive increase of inhomogeneity in the atom interior 
due to the inter-shell transition,
characterized by the growth in the region of $(s^2,q)$ parameter space probed by
each successive system. 
In Figs.~\ref{fig:vxNe} and~\ref{fig:vxBe} we show the logarithm of the density,
$s^2$, and the absolute values of $3q$ and $3x$ in (a); 
we show exchange potentials in (b).

In each case, the Laplacian-derived parameter 
$3q$ 
is a reliable indicator of shell structure --
it is negative at the nuclear cusp and at the peak of the valence shell 
(shown as the dashed portions of the curve in each figure)
and positive in the transitional region between shells and asymptotically.  
The transition between shells 
is also indicated by an abrupt change in 
slope in $ln(n)$,  
which coincides with a maximum in $\lapln$ and $3q$.  
The gradient-derived $s^2$ is less sensitive 
to structural details, but does exhibit an oscillation at the core-valence
boundary.  The hybrid parameter $3x$ retains the negative singularity at the cusp 
exhibited by $3q$ but a has much diminished region of negative value at the valence shell 
peak, which vanishes altogether for Ne.

With a completely filled valence shell, the transition between the core and 
valence densities in Ne is fairly gradual
and as a consequence, the inhomogeneity parameters $s^2$, $3q$, 
and $3x$ are not too far from the GEA limit in the inter-shell region. 
For Be, a combination of low net valence charge and a large change in 
the decay rate of the density makes the core-valence transition one of severe 
inhomogeneity, 
with $3q$ topping a value of 24 and $s^2$ a value of 8, both
well beyond the gradient expansion criterion of $3x\!\ll\!1$.  
Thus Be is a particularly good stress test for the behavior of density 
functionals with respect to severely rapid change in density.

Exchange potentials for the DFT models considered in Table~\ref{table:energy} 
are shown for Ne in Fig.~\ref{fig:vxNe}(b) 
and Be in Fig.~\ref{fig:vxBe}(b).  
The exact potential, like in He, proves hard to match -- 
it lies consistently below that of all DFT potentials, and 
has an oscillation at the core-valence transition not captured by the
DFT models.
The LDA faithfully follows the general trend of this potential
but, as for He, shifted upwards by roughly a constant amount.  
The mGEA gives
the best qualitative fit of the inter-shell region,
at the price of 
catastrophic failure at large $r$.  
The imposition of a bound on the X functional for the large-inhomogeneity 
limit in the APBE and the two mod-GGA's overcorrects the 
GEA potential in the inter-shell region, leading to a less accurate result here. 

The effects of increasing inhomogeneity as one goes from He to Ne to Be
bring about some surprising results, particularly for the SOGGA-q model
that depends solely on $q$, the Laplacian-derived GEA variable.  
The moderately large inhomogeneity encountered in Ne at the core-valence
boundary induces oscillatory behavior in the SOGGA-q model that is 
significantly larger than that of the APBE or modAPBE.  This does
indicate that the Laplacian is more sensitive to structural details than
the gradient of the density 
-- unfortunately the change in potential is in the wrong direction.  
A more unpleasant surprise occurs with the more inhomogeneous Be atom:
huge oscillations appear in the potential, associated with the change in 
sign in $q$ at $r\!=\!1.4$.  
Apparently the potential has a nonlinear sensitivity to rapid 
changes in $q$ that was not apparent for the He atom.

The mGEA, APBE and modAPBE also show a similar, if less dramatic
dependence on the inhomogeneity in the transition region, with 
trends for Ne for each case exaggerated in Be.
In the more homogeneous Ne atom, there is 
very little difference between the three models, 
a reflection of 
the relative closeness between the variables $s^2$ and $3x$, 
and their lying largely within the gradient expansion limit.
The differences between models are larger for Be, and lead
to larger errors with respect to the exact potential.
The additional oscillatory structure, genuine or spurious, in the potentials
for Ne and Be show up in the curvature $I$ shown in 
Table~\ref{table:energy}.  Those for Ne are not much larger than for He, but
the considerable additional structure for Be causes an order of magnitude
jump in this integral.

\begin{figure}
\includegraphics[clip=true,width=0.40\textwidth]{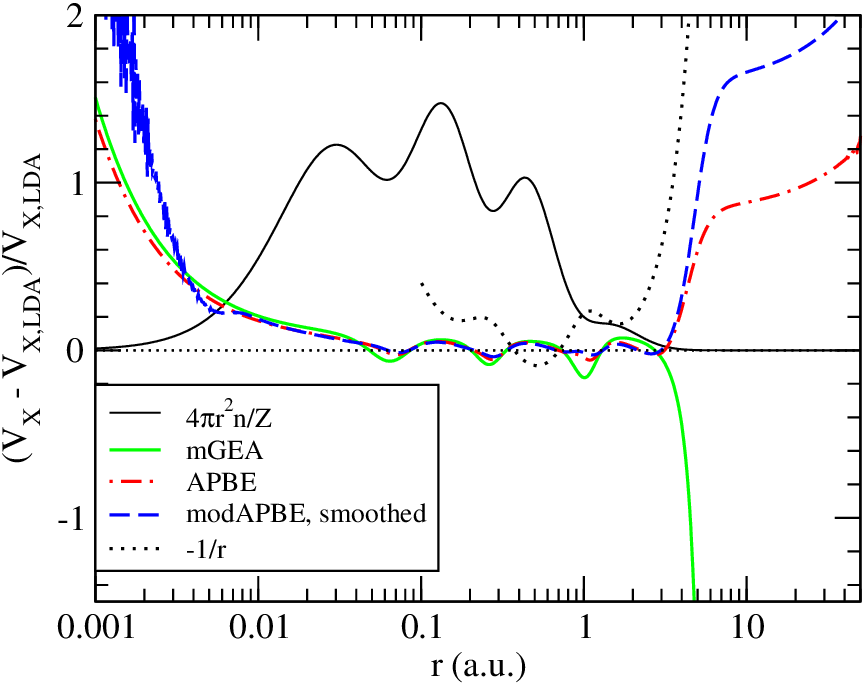}
\caption{\label{fig:KrGEA}
(color online)
X potentials for the Kr atom, obtained for the APBE density using the APE pseudopotential generator.
Green solid lines shows the mGEA model, 
red dot-dashed line shows the APBE using the conventional $s^2$ GEA parameter,
blue dashed line shows the modAPBE using the hybrid GEA parameter $x$,
black dotted line shows asymptotic behavior of true exchange potential.
Noise in modAPBE data is due to round-off error.
An arbitrarily scaled radial density profile (black solid) is included as a 
guide to the eye.
}
\end{figure}

\subsection{Large atoms}

The exchange potential for a typical larger-$Z$ closed-shell atom, Krypton,
is shown in Fig.~\ref{fig:KrGEA}.  This figure shows
the relative difference between several gradient-expansion-derived exchange potentials 
and that of the LDA.
These are calculated from a standard all-electron calculation using the 
APE pseudopotential generator,~\citep{APE} using the APBE model of exchange 
and correlation.  As a guide to interpretation, the radial probability
density for Kr is shown in black at an arbitrary scale, showing clearly the 
four filled shells of this atom.

All the gradient-corrected potentials are very close to each other, and to 
the LDA, in the Thomas-Fermi scaling region -- for distances outside the 
nuclear cusp $r\!>\!a_0/Z$ up to the asymptotic edge of the valence shell 
at $r\!\sim\! 4~a_0$.  
In this region, where the large proportion of electron density
is located, the GEA is a good approximation and there is no 
appreciable difference
between GEA, GGA or modGGA.  Three dips in the GEA curve, indicating 
a slightly larger disagreement with the LDA, mark the transitions 
between the four energy shells and thus intermittent departures
from pure scaling behavior.  

Plotted as a relative
difference with the LDA, the correct limiting behavior of the
X potential of $-1/r$ shows up as 
a positive exponential curve, which is shown for large $r$ as a dotted line.  
Note 
that it produces a roughly constant shift with the LDA for intermediate 
values of $r$ where the Thomas-Fermi approximation is valid.  

The mGEA has a singularity at the nuclear cusp 
independent of $\alpha$ 
(in the case of the $\alpha\!=\!1$ or $q$-only limit, this is 
because the GEA lacks the $1/q$ limiting behavior needed to produce a 
finite potential.)  
It also diverges from the LDA value asymptotically in the classically forbidden region 
of the atom, which like the cusp is not treatable in
the semiclassical Thomas-Fermi approach.  Here it becomes nonzero and 
diverges exponentially to $+\infty.$
The APBE is identical to the mGEA in the cusp region, tamps down on its oscillatory 
behavior between shells and has a reasonable long-range behavior.  
The modAPBE does slightly better than the APBE in the asymptotic limit,
in comparison to the correct power-law behavior, because of the larger
value of $\kappa$ used.
It is hard to qualify it as better in the cusp regime because of 
excessive numerical noise that appears in taking numerical
derivatives for $\nabla^2 n$ and $V_X$.
This noise 
may be reduced by increasing the precision of the 
the numerical density data used to generate the potential, typically to 
fifteen significant figures. 
Nevertheless, it points to a difficulty in implementing our approach -- 
numerical techniques adequate for gradient-based models may
not work for Laplacian-based models without retuning.~\cite{JemmerKnowles}

\section{Discussion\label{sec:discussion}}
\subsection{Future steps}
Our functionals are currently limited by a lack of significant 
information beyond that of the GGA.  
They are identical 
to the GGA in the gradient expansion limit
and have the same scaling constraints. The large inhomogeneity
limit is determined by a fit to the GGA for low-$Z$ atoms. 
Since they encode the same physics, it is not surprising that our new 
models closely match the GGA in calculated energies and potentials.  
The significant difference is tied to the genuinely new piece of physics 
we have introduced -- the behavior of the exchange hole near
the nucleus, and thus the behavior of the exchange potential in this region.

It is possible to do better.  A next step in DFT development,
taken normally in the development of meta-GGA's, is to satisfy the 
next-highest or fourth-order term in the gradient expansion for exchange. 
This involves Laplacians and gradients of the density in a way that cannot
be trivially unentangled by an integration by parts.
Ignoring the issue of whether that expansion is better derived 
from the extended Thomas-Fermi theory of the atom or from the homogenous electron 
gas, the fourth-order gradient expansion about the latter is:~\citep{Svendsen}
\be
  F^{GEA}_X(s^2,q) = 1 + \frac{10}{81} 3x + \frac{146}{2025}q^2 + \frac{73}{405}s^2q + Ds^4
\ee
where $D$ is believed to be 0, and $3x$ is the arbitrary linear combination of
$s^2$ and $q$ of Eq.~(\ref{eq:x}).  
As we already use the explicit linear combination of $s^2$ and $q$ to describe the 
lowest order correction of the gradient expansion, we automatically generate terms of 
the correct sort for the higher-order corrections as well.  In fact, if we take 
the choice $3x\!=\!0.6q \!+\! 0.8s^2$ derived from the gradient expansion of the X hole
we can come remarkably close to the fourth-order correction with
\be
  F^{modGEA}_X(x) = 1 + \frac{10}{81} 3x + \frac{1}{5}(3x)^2,
\ee
with a difference $\Delta F_X$ from the exact value of 
\be
   \Delta F_X(s^2,q) = -0.0001 q^2 + 0.012s^2q + 0.128s^4.
\ee
Here 
the $s^2q$ coefficient is within 10\%
of the exact value; the 
nonzero value for $D$ is close to that used in a previous metaGGA.~\cite{PKZB}
This nice result is all the more surprising in that the arguments
we have used in generating our choice for $x$ are strictly related to the lowest order
in perturbation theory.  

In our current form for $F_X$ [Eq.~(\ref{eq:fxmodGGA})] there is a free 
parameter $\eta$ that
determines the coefficient of $x^2$; however it is already used to minimize the 
curvature of the derivative of the exchange energy density with $\lapln$, and
cannot also be used to match the correct fourth-order correction.  In fact, the 
curvature minimization requirement produces a fourth-order gradient 
correction with the wrong sign, and thus wrong qualitative results.  
This may be the reason
why our X potential seems to be slightly worse than 
the GGA in the core-valence transition region of Be as shown in 
Fig.~\ref{fig:vxBe}.

A necessary further step will be to construct a correlation functional to match
the exchange functional introduced here.  The goal is to model a structure of the 
form~\cite{PBE}
\be
   H_C \sim \log\{1 + (\beta/\gamma) t^2 F_C[A(r_s)t^2]\}
\ee
where $F_C$ is an enhancement factor similar to that used for exchange, 
$A(r_s)$ a function and $\gamma$ a parameter both fixed by scaling constraints, 
$\beta$ defines the strength of the gradient correction in the
low inhomogeneity limits, and $t^2 \sim s^2/r_s$ is the measure of 
inhomogeneity given by Eq.~(\ref{eq:tsq}).  
The minimum step needed to generate a useful correlation 
functional is to replace $s^2$ with the regularized hybrid variable 
$3\xbar$ and use variational 
techniques to control the resulting potential.  
A complicating issue is the stringent constraint necessary to have 
a physically reasonable energy density -- the argument to
the logarithm must be greater than zero, which is harder
to achieve for negative $x$ than the absence of a pole in $F_C$.
Secondly constraints that would add value to the functional beyond
that of the GGA are unknown.  What is nature of correlation potential or energy
in the limit of large negative $x$, i.e. at the nucleus?  What is the correct
response to the fourth order gradient expansion for exchange?  Do problems
with gauge-variant constraints occur -- e.g. the limit of large $t^2$?

\subsection{Conclusions\label{sec:conclusions}}
The major findings of this project have been the greater understanding of the 
physics behind the generalized gradient approximation of DFT.  We have shown 
that in principle, a GGA good for a large range of systems and conditions can 
be constructed starting from any linear combination of $\gradnsq$ and $\lapln$
that produces the same gradient expansion correction, giving rise to an 
infinite family of ``gauge choices" parameterized by a linear coefficient $\alpha$, from $\alpha\!=\!0$, or gradient only, 
to $\alpha\!=\!1$, or Laplacian only.  This model then fits all 
constraints that any standard GGA fits, as long as they can be framed
in a specifically gauge-invariant way. 
This result gives the DFT development community a greater degree 
of flexibility in constructing future DFT's.

We have found some fundamental differences between gauge choices, however, 
and with them, some important caveats to the finding above.  
The two extremes $\alpha\!=\!0$ and $\alpha\!=\!1$ each have important 
defects which lead to failure when the result of the gradient expansion limit 
is extended to all density-functional space.  
The gradient-only limit fails at the nucleus, with a spurious singularity
in the potential, but this problem likely has a small effect for 
most atoms.  
The Laplacian-only limit is plagued by spurious oscillations in the potential
that we have been unable to control to a reasonable degree.
And these occur in places,
such as the valence shell of atoms, or the transition between shells, where
these problems cannot be ignored.
Thus the $\alpha\!=\!1$ 
case that we have explored must be considered the worse choice.

However hybrids work.  Just like hybrids in other areas of science, 
or, for that matter, in other areas of density functional theory,
the combination of two alternate formulations of a problem lead to 
formulation that is superior to both.   A choice of $\alpha\!=\!0.2$,
inspired by a gradient expansion of the exchange hole, cures the problems
of both conventional GGA and our exploratory Laplacian-based approach,
thus providing the best of both worlds.  This supports 
the philosophy of DFT development based on modeling the XC hole that 
was the original inspiration for this work -- it is in paying attention
to the XC hole that an optimal hybrid solution is derived.

The technique of eliminating false oscillations in the exchange potential 
by the minimization of the curvature of $\partial \exc/\partial \lapln$ is a useful
technical advance.  It is the key here to finding a successful 
functional form that includes $\lapln$ and produces consistently stable and
reliable potentials over a wide variety of systems.
It may seem that our hybrid-variable functional is so close a match to the 
GGA because it is a trivial extension of it.  But this obscures the fact that 
it would certainly have performed worse
without the optimization of variational parameters
$\eta$ and $C$.  And the final choices made for these parameters were very much 
unexpected.  In other words, it is only after a ``hidden constraint" that the
X potential should be as free of unphysical curvature as possible that the connection 
between the Laplacian-based GGA and conventional GGA becomes evident.
Such a technique should prove useful in future attempts to 
produce workable models for orbital-free XC or kinetic energy densities. 

Not all constraints are what they seem.  
The local Lieb-Oxford bound that is a key
component of conventional GGA's has had to be
rethought.  As a constraint on the 
energy density, and not the energy, it is not a physical constraint and must be 
altered if one changes the ``gauge" of the energy density, say by shifting from 
density gradient to density Laplacian.  The global Lieb-Oxford bound is a true
physical constraint, but is a bound and a very loose one in practice, and 
thus of less value than other constraints 
in DFT such as scaling laws and limit cases which are exact conditions. 
An exact physical constraint equivalent to the local Lieb-Oxford bound, 
one that is generally applicable and exact, 
is the low-$Z$ limit of atomic exchange energies, with the contrasting
constraint of weak inhomogeneity provided by the large-$Z$ limit.
This approach matches the LDA to one set of exact physical results (the
homogeneous electron gas) and the GGA to another (atomic energies).
It reflects the relative importance of the contribution of 
``surface" of the atom, which requires a GGA correction, to the 
``volume" where extended Thomas-Fermi theory and hence the gradient expansion
is exact.

Much of what one would 
like to fix in the GGA X potential for atoms -- bad treatment of the oscillations
at shell boundaries and failure to handle the $1/r$ asymptotic behavior of the
true potential, is not fixed by our models.
This is due in large part to the essential limitation of our approach as an 
alternative ``gauge" for the GEA: many energetically relevant regions of 
electronic systems are near the gradient expansion limit, where there is with 
mathematical certitude no difference between our models and conventional GGA's.  
Secondly, where we are not at this limit, it is not obvious that the 
Laplacian can provide the necessary information to fix the GGA.
Semilocal models, those involving only the local density and its derivatives,
have the limitation of being tied to 
the local environment, while real systems, especially in such important cases as 
the covalent bond, have inherently nonlocal aspects to them.  Thus, for example,
the inability of all semilocal models to obtain the correct $1/r$ limit of the 
X potential seems to be inherently a property of self-interaction error that may 
not be removable except at a higher level of theory.
Ultimately, it seems that it is the constraints imposed, 
once they are defined properly, and once hidden ones such as curvature 
minimization are identified, that define the performance of 
the DFT models we have studied.  Our model is a net improvement over the GGA 
not so much because it uses the Laplacian of the density but because it 
fits constraints at the nucleus that are not satisfiable by the GGA. 

Within this context, there is some hope for improvement.  
The transition region between atomic shells, one of relatively high 
inhomogeneity, is qualitatively poorly described by GGA potentials, and our 
more general model does not help.
However, this may be due to some degree to the fact that 
it does not capture the proper fourth-order gradient expansion 
for the exchange energy.  This is quite within the capabilities
of our approach to DFT, by generalizing the
enhancement factor $F_X(x)$ to have the correct $x^2$ coefficient while
otherwise keeping curvature in the potential to a minimum.
A second opportunity is to explore areas of electronic topology that are not 
accessible with atomic densities.  The most important such case would be the chemical 
bond, particularly the situation of large positive $q$ and
zero $s^2$ that occurs in a bond as its constituent atoms become dissociated.  
(Covalent bonds explore negative but small values of $q$ that presumably could
be mapped back to a GGA.)  The information about electronic topology 
that the Laplacian could bring to the case of bond dissociation 
may help improve DFT
predictions of bondlengths and molecular potential surfaces.


\begin{acknowledgments}
One of us (ACC) thanks Cyrus Umrigar for useful discussions and for providing 
atomic density and potential data used in this paper, and Neal Coleman 
and Shaun Wood for help in writing numerical code.
Work supported by National Science Foundation grant DMR-0812195.
\end{acknowledgments}


%


\begin{thebibliography}{78}
\bibitem [{foo()}]{footnote1}%
  \BibitemOpen
  \href@noop {} {}\bibinfo {note} {Any scaling law can be framed in terms of
  the scaling of the integrated XC energy, and is thus naturally gauge
  invariant.}\BibitemShut {Stop}%
\bibitem [{LOF()}]{LOFootnote}%
  \BibitemOpen
  \href@noop {} {}\bibinfo {note} {The correlation energy density of the PBE
  tends to zero in the limit of high inhomogeneity and thus drops out of
  equation.}\BibitemShut {Stop}%
\bibitem [{Cus()}]{CuspFootnote}%
  \BibitemOpen
  \href@noop {} {}\bibinfo {note} {The solution used in the recent TPSS
  metaGGA~\cite{TPSS} is to require $a(s^2)$ have zero slope at the value $s^2
  = 0.18$ typical of the nuclear cusp of most atoms, when the kinetic energy
  density $\tau$ approaches the von-Weizacker value that characterizes the 1s
  shell. It is thus a special case of equation Eq.~(\ref{eq:fxcusp}) with
  $b=0$.}\BibitemShut {Stop}%
\end{thebibliography}
\end{document}